\documentclass[galaxies,article,accept,moreauthors,LaTeX]{Definitions/mdpi}

\firstpage{1}
\pubvolume{xx}
\issuenum{1}
\articlenumber{5}
\pubyear{2020}
\copyrightyear{2020}
\history{Received: date; Accepted: date; Published: date}
\usepackage{hyperref}
\usepackage{upgreek}
\usepackage{pdflscape}
\Title{Asteroseismic Analysis of $\delta$~Scuti Components of Binary systems: The Case of KIC~8504570}

\Author{Alexios Liakos $^{1,}$* \orcidA{} and Panagiotis Niarchos $^{2}$}

\AuthorNames{Alexios Liakos and Panagiotis Niarchos}

\address{%
$^{1}$ \quad Institute for Astronomy, Astrophysics, Space Applications and Remote Sensing, National Observatory of Athens, Metaxa \& Vas. Pavlou St., GR-15236, Penteli, Athens, Greece; alliakos@noa.gr\\
$^{2}$ \quad Section of Astrophysics, Astronomy and Mechanics, Department of Physics, National and Kapodistrian University of Athens, GR-15784, Zografos, Athens, Greece; pniarcho@phys.uoa.gr}

\corres{Correspondence: alliakos@noa.gr}%; Tel.: (optional; include country code; if there are multiple corresponding authors, add author initials) +xx-xxxx-xxx-xxxx (F.L.)}

\abstract{The present work concerns the Asteroseismology of the $Kepler$ detached eclipsing binary KIC~8504570. Particularly, it focuses on the pulsational behaviour of the oscillating component of this system and the estimation of its physical parameters in order to enrich the so far poor sample of this kind of systems. Using spectroscopic observations, the spectral type of the primary component was determined and used to create accurate light curves models and estimate its absolute parameters. The light curve residuals were subsequently analysed using Fourier transformation techniques to obtain the pulsation models. Theoretical models of $\delta$~Scuti stars were employed to identify the oscillation modes of the six detected independent frequencies of the pulsator. In addition, more than 385 combination frequencies were also detected. The absolute and the pulsational properties of the $\delta$~Scuti star of this system are discussed and compared with all the currently known similar cases. Moreover, using a recent (empirical) luminosity-pulsation period relationship for $\delta$~Scuti stars, the distance of the system was estimated.}

  % List three to ten pertinent keywords specific to the article, yet reasonably common within the subject discipline.

\begin{document}
%%%%%%%%%%%%%%%%%%%%%%%%%%%%%%%%%%%%%%%%%%

%%%%%%%%%%%%%%%%%%%%%%%%%%%%%%%%%%%%%%%%%%
\section{Introduction}
\label{sec:intro}
%%%%%%%%%%%%%%%%%%%%%%%%%%%%%%%%%delta Sct General
The $\delta$~Scuti stars are short-period and multiperiodic pulsating variables. In general, they oscillate in radial and low-order non-radial pulsations due to $\kappa$-mechanism \citep[c.f.][]{AER10, BAL15}. However, recently, it has been proposed that the turbulent pressure in the Hydrogen convective zone may explain the observed high-order non-radial modes \citep{ANT14, GRA15}. Their masses typically range between 1.4-2.5~$M_{\odot}$ \citep{AER10}, their spectral types between AIII/V-FIII/V, and they are located inside the classical instability strip. Thanks to the $Kepler$ \citep{BOR10, KOC10}, the $K2$ \citep{HOW14}, the GAIA \citep{GAI16}, and the Transiting Exoplanet Survey Satellite \citep[TESS;][]{RIC09, RIC15} missions, as well as the All-Sky Automated Survey for Supernovae \citep[ASAS-SN;][]{SHA14} project, many research works \citep[e.g.][]{BOW18, MUR19, ZIA19, BED20, JAY20}, based on large data sets of $\delta$~Scuti stars, have been published providing new tremendous knowledge for this kind of pulsators.

%%%%%%%%%%%%%%%%%%%%%%%%%%%%%%%%%%%%%%%EBs General
The eclipsing binaries (EBs) can be considered as the utmost tools for the calculation of stellar absolute parameters (e.g. masses, radii, luminosities) and the evolutionary stages of their components, particularly in case when spectroscopy and photometry are combined. However, it should be noted that for systems with large luminosity difference between their components, the radial velocity measurements of the less luminous component is, in general, a very hard task because the light of more luminous component dominates the spectrum. 

In addition, given that the phase parts along the quadratures are the most important for the calculation of the amplitudes of the radial velocity curves, the systems with orbital periods longer than 1-2 days need observations in different and, depending on the orbital period, possibly distant dates. Moreover, particularly for the $Kepler$ systems (with a luminosity range of approx. 10-15 mag) at least 2-4~m size telescopes have to be employed. Hence, according to these limitations, telescope time is not easy to be allocated. Therefore, for the aforementioned reasons, in the absence of radial velocity curves, the least-squares minimization technique has to be applied to the photometric data in order to estimate the parameters of the components. Moreover, another powerful tool of the EBs is the `Eclipse Timing Variations' (ETV) method, which allows to detect mechanisms (e.g. mass transfer, tertiary component etc.; c.f. \citet{BUD07}-ch.8 and \citet{BOR16} and references therein) that modulate the orbital period.

%%%%%%%%%%%%%%%%%%%%%%%%%%%%%%%%%%%%%%% pulsating stars in EBs
Specifically, the subject of $\delta$~Scuti stars in EBs is extremely interesting because it combines two totally different topics of Astrophysics and provides the means for remarkable results. On one hand, these systems host an oscillating component, whose pulsational properties are directly measurable by analysing photometric/spectroscopic data. On the other hand, the geometric phenomena of eclipses, that occur during the orbital cycles, can be used to determine the absolute properties of the components of these systems. Therefore, the study of this kind of systems, especially the detached ones with wide orbits, allows for the direct determination of the physical properties of pulsating stars. The latter can be further used to correlate their pulsation properties with their evolutionary stage and, in the future, to constrain further the current evolutionary models of the pulsating stars. Furthermore, the close detached eclipsing systems (i.e. with orbital periods of the order of a few tens of days) and the semi-detached binaries have opened a new window for examining the influences of the binarity and the mass transfer on the pulsations.

%%%%%%%%%%%%%%%%%%%%%%%%%%%%%%%%%%%%%%% History of pulsating stars in EBs
Approximately 20 years ago, \citet{MKR02} suggested the term `$oEA~stars$' (oscillating eclipsing binaries of Algol type) for categorizing the EBs with a $\delta$~Scuti mass accretor component of (B)A-F spectral type. A few years later, the first connection between orbital ($P_{\rm orb}$) and dominant pulsation ($P_{\rm pul}$) periods for these systems was published by \citet{SOY06a}. \citet{LIA12} performed a long-scale observational survey on more than 100 candidate systems and resulted in a publication of a catalogue with 74 cases and updated correlations between fundamental parameters for these systems. The first try for theoretical justification for the $P_{\rm pul}-P_{\rm orb}$ correlation was made by \citet{ZHA13}, who derived a similar empirical relation with the previous studies, based on different assumptions (i.e. the coefficient of $\log P_{\rm orb}$ is 1), and found that its slope may be a function of the pulsation constant, the filling factor of the oscillating component and the mass ratio of the binary system. \citet{LIAN15, LIAN16} announced the existence of a possible boundary in the $P_{\rm orb}$ ($\sim13$~d) beyond that $P_{\rm pul}$ and $P_{\rm orb}$ can be considered uncorrelated. \citet{KAH17}, based only on eclipsing systems, suggested an almost doubled value for this boundary. \citet{LIAN17} published the most coherent catalogue for these systems to date (available online\footnote{\url{http://alexiosliakos.weebly.com/catalogue.html}}), providing also updated correlations between the fundamental parameters of these systems distinguished according to their geometrical status and the reliability of their absolute parameters. An extended review for binaries with pulsating components was published by \citet{MUR18b}. \citet{MUR18}, based on a sample of 2224 $Kepler$ $\delta$~Scuti light curves, employed the pulsations timing technique \citep{SHI12, SHI15, MUS15} and identified 341 new binaries with long $P_{\rm orb}$ ($>100$~d) that host a $\delta$~Scuti component. \citet{LIA20}, using recently discovered systems, presented updated correlations between $P_{\rm pul}-P_{\rm orb}$ and $P_{\rm pul}-\log g$ for close (i.e. $P_{\rm orb}<13$~d) detached eclipsing binaries with a $\delta$~Scuti component.

%%%%($Kepler$, Binaries, accuracy, no gaps for asteroseismology, database),
The data quality of both $Kepler$ and $K2$ missions provided new insights for Asteroseismology. Due to their unprecedented accuracy (i.e. order of a tens of mmag), they allow the detections of low amplitude frequencies of the order of a few $\upmu$mag \citep{MUR13}. Moreover, the continuous data acquisition from these missions for relatively long periods of time practically extinguishes the alias effect \citep{BRE00} in frequencies detections. Especially the time resolution of the short-cadence data ($\sim1$~min) has been proven as extremely useful for the studies of short-period pulsating stars, such as the $\delta$~Scuti stars. Furthermore, the data of these missions have been widely used for the study of EBs. Specifically for the latter systems, an excellent online catalogue, namely `$Kepler$ Eclipsing Binary Catalog' \citep[$KEBC$,][]{PRS11}, that is publicly available\footnote{\url{http://keplerebs.villanova.edu/}}, has been created and includes the detrended data and other useful information for a few thousands of EBs.

%%%%%%%%%%%%%%%%%%%%%%%%%%%% Motivation
The present work is a series paper on individual EBs with a $\delta$~Scuti component \citep[see also][]{LIA09, LIA13, LIA14, LIAN16, LIA17, LIA18, LIA20}. The system KIC~8504570 was selected for the present work because its pulsational behaviour is currently unknown to the community and its detailed analysis contributes to the sample of $Kepler$ detached binaries with a $\delta$~Scuti member (27 systems in total published to date; see \citet{LIAN17} and updated lists in \citet{LIA20}).

%%%%%%system
KIC~8504570 (2MASS~J19405685+4430276) was discovered by the $Kepler$ mission \citep{SLA11} and has an orbital period of $\sim4$~d. The only existing references concern mostly its temperature determination \citep[6874-7390~K;][]{SLA11, PIN12, ARM14, HUB14, FRA16, QIA18, BER18, BAI19}, while \citet{DAV16} included it in the list of $Kepler$ systems that present flare activity.

%%%%%%%%%%%%%%%%%%%%%%%%%%%% Outline + tips
Details about the ground-based spectroscopic observations as well as the estimation of the spectral type of the primary component of the system are given in Section~\ref{sec:sp}. The $Kepler$ light curves (LCs) analyses, the modelling results, and the absolute parameters calculation are presented in Section~\ref{sec:LCmdl}. Section~\ref{sec:Fmdl} includes the frequency search on the LCs residuals, the pulsation models, and the oscillations modes identification. Finally, Section~\ref{sec:Dis} contains the summary of this work, a comparison in terms of evolution and properties of this system with other similar cases, discussion, conclusions, and future prospects.

%%%%%%%%%%%%%%%%%%%%%%%%%%%%%%%%%%%%%%%%%%%%%%%%%%%%%%%%%%%%%%%%%%%%%%%%%%%%%%%%%%%% S E C T I O N 2 ----- SPECTROSCOPY
%%%%%%%%%%%%%%%%%%%%%%%%%%%%%%%%%%%%%%%%%%%%%%%%%%%%%%%%%%%%%%%%%%%%%%%%%%%%%%%%%%%%
%%%%%%%%%%%%%%%%%%%%%%%%%%%%%%%%%%%%%%%%%%%%%%%%%%%%%%%%%%%%%%%%%%%%%%%%%%%%%%%%%%%%
%%%%%%%%%%%%%%%%%%%%%%%%%%%%%%%%%%%%%%%%%%%%%%%%%%%%%%%%%%%%%%%%%%%%%%%%%%%%%%%%%%%%

\section{Spectroscopy}
\label{sec:sp}

%%%%%%% ATS, observations + data reduction
The purpose of the spectroscopic observations was the estimation of the spectral type of the primary component of the system. The spectra of the target were obtained with the 2.3~m Ritchey-Cretien `Aristarchos' telescope at Helmos Observatory in Greece on 6th October 2016. The \emph{Aristarchos~Transient~Spectrometer}\footnote{\url{http://helmos.astro.noa.gr/ats.html}} (ATS) instrument \citep{BOU04} using the low resolution grating (600~lines~mm$^{-1}$) was employed for the observations. This set-up provided a resolution of $\sim3.2$~{\AA}~pixel$^{-1}$ and a spectral coverage between approximately 4000-7260~{\AA}. Three successive spectra with 10~min exposures were acquired for KIC~8504570 during the orbital phase 0.81 and added together in order to achieve a better signal-to-noise ratio (S/N). The mean S/N of the individual spectra was $\sim13$, while that of the final integrated spectrum was $\sim18$. For the spectral classification, a spectral line correlation technique between the spectrum of the variable and standard stars was applied. The selected standard stars, suggested by the Gemini Observatory\footnote{\url{https://www.gemini.edu/observing/resources/near-ir-resources/spectroscopy}}, ranged between A0-K8 spectral classes (one standard star per subclass) and were observed with the same set-up during August-October 2016. All spectra were calibrated (bias, dark, flat-field corrections) using the \textsc{MaxIm DL} software. The data reduction (wavelength calibration, cosmic rays removal, spectra normalization, sky background removal) was made with the \textsc{RaVeRe} v.2.2c software \citep{NEL09}.

%%%%%%%%%% Spectral classification method in brief
The applied correlation method has been described in detail in \citet{LIA17}, but is briefly presented also here. For the comparison between the spectrum of KIC~8504570 and the standard stars, the Balmer and the strong metallic lines between 4000-6800~{\AA} were used. The differences of spectral line depths between each standard star and the target star derive sums of squared residuals in each case, with the least squares sums to indicate the best fit. This method is quite efficient in cases of EBs with large luminosity difference between their components, because the total spectrum is practically dominated by the light of the primary star. In our study we did not use any synthetic spectrum approximation \citep[c.f.][]{JOS17} in order to avoid any instrumental effects (e.g. distortion) that cannot be taken into account in a synthetic model. Therefore, using the direct comparison method, and given that all spectra were taken with the same set-up, any systematic effects are directly vanished (i.e. differential removal of systematics).

On the other hand, in cases with small luminosity difference between the components, this method does not provide accurate results, and more specifically, it might lead to an underestimation of the spectral type of the primary. Therefore, in order to avoid this, the following method \citep[described also in][]{LIA20} was applied. Using the spectra of the standard stars, all the possible combinations were calculated by simply adding and normalizing the spectra. Furthermore, for every spectra combination the spectrum of each component was given weight between 0-1 denoting its light contribution to the combined spectrum. The starting value for the contribution of the primary component was 0.5 and the step was 0.05. Finally, for each spectra combination, ten sub-combinations with different light contributions of the components were derived. Similarly to the previous method, the same spectral lines were used for the comparison of the combined spectra with that of KIC~8504570, deriving again sums of squared residuals. Hence, the least value of these residuals indicated again the best match.

\begin{figure}[h]
	\centering
	\includegraphics[width=13cm]{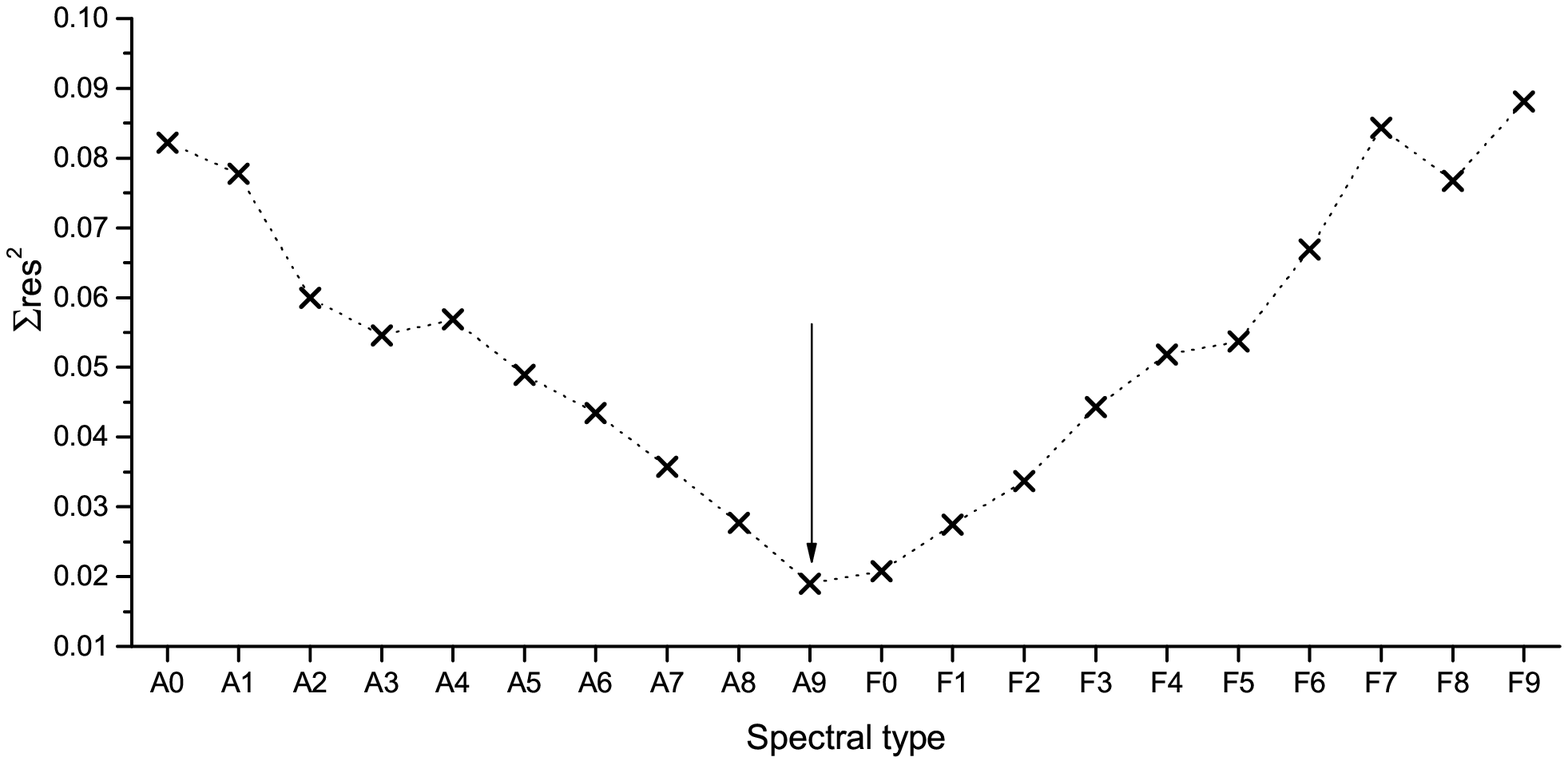}
	\caption{Spectral type-search plot for KIC~8504570. The arrow indicates the adopted spectral type for the primary component. The comparison is shown only between A0-F9 spectral types due to scaling reasons.}
	\label{fig:STs}
	%\end{figure}
	%\begin{figure}
	\includegraphics[width=13cm]{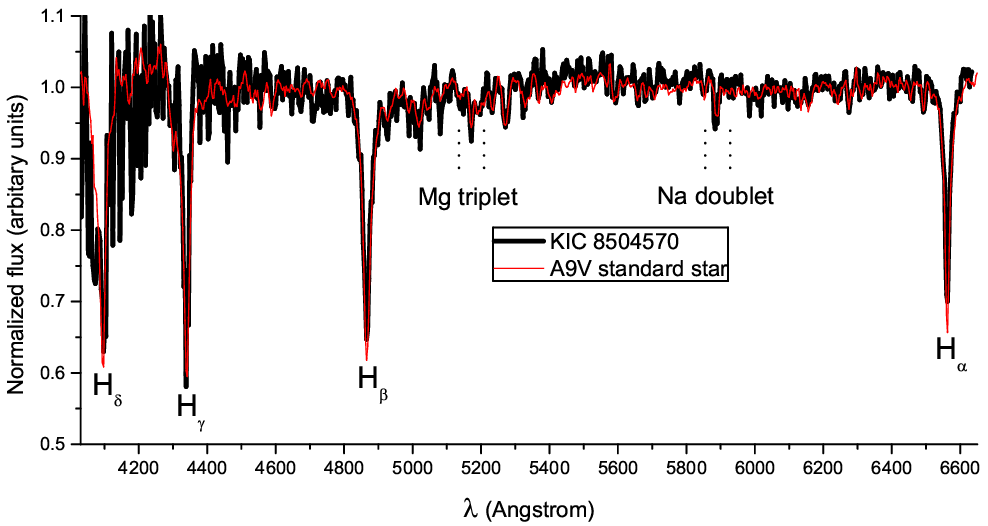}\\
	\caption{Comparison spectra of KIC~8504570 (black line) and the standard star (red line) with the closest spectral types. The Balmer and some strong metallic lines are also indicated.}
	\label{fig:spectra}
\end{figure}

%%%%%%%%%%%%%%%%%%results
The spectrum of KIC~8504570 was found to be dominated by at least 95\% by the light of the primary component. Therefore, its spectrum was directly compared with those of the standards. The sum of squared residuals against the spectral type for this system is plotted in Figure~\ref{fig:STs}, which shows that the best fit was found with the spectrum of an A9V standard star. The spectrum of the system along with that of best-match standard star are illustrated in Figure~\ref{fig:spectra}. It should be noted that for the spectra continuum normalization, polynomials of various orders were used according to the spectral types of the stars, since each spectral type has different peak wavelength. However, for the continuum normalization of the spectra of the standard stars with spectral types close to that of the target (e.g. between A5-F5) the same polynomials were used. Therefore, since our method is based on direct comparison (i.e. subtraction of spectra), the non-perfect continuum normalization does not affect the results. The present spectral classification, with an error assumption of one sub-class, corresponds to a temperature $T_{\rm eff}=7450\pm150$~K for the primary, based on the relations between $T_{\rm eff}$ and spectral types of \citet{COX00}. The present result comes in relatively good agreement with those given in previous studies (see Section~\ref{sec:intro}).

\section{Light curves modelling and absolute parameters calculation}
\label{sec:LCmdl}
%%%%%%%%%%%%%%%% data and selection

\begin{figure}[t]
	\centering
	\includegraphics[width=\columnwidth]{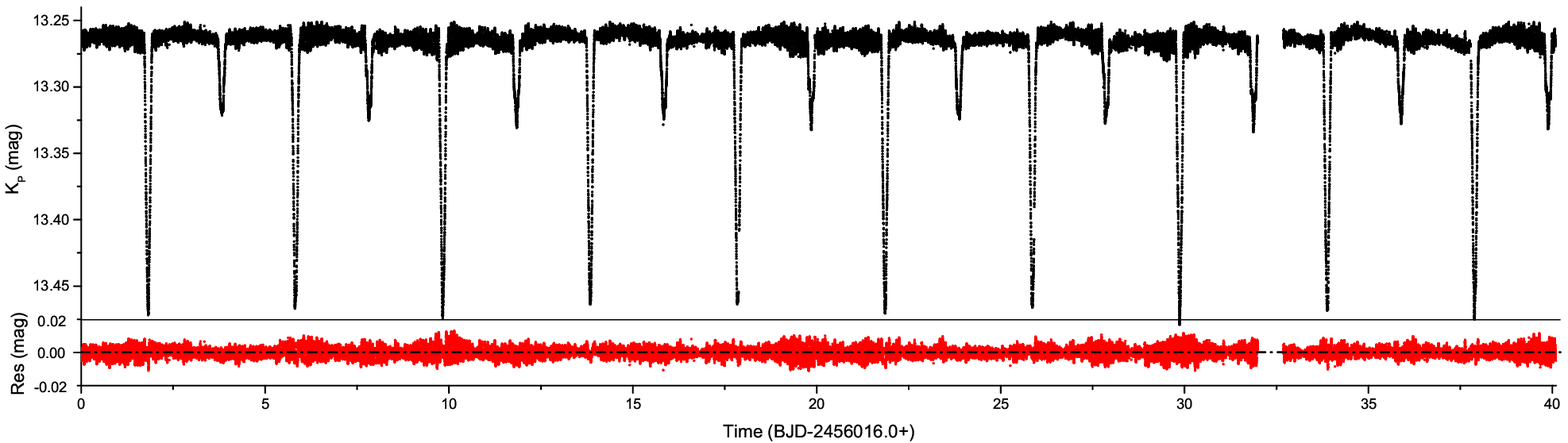}
	\caption{Short-cadence LCs (black points) for KIC~8504570 and their residuals (red points) after the subtraction of the LC models. The plotted data concern only the first 40 days of observations, while the rest are not shown due to scaling reasons.}
	\label{fig:LCsandRes}
	%\end{figure}
	%\begin{figure}
	\centering
	\includegraphics[width=12cm]{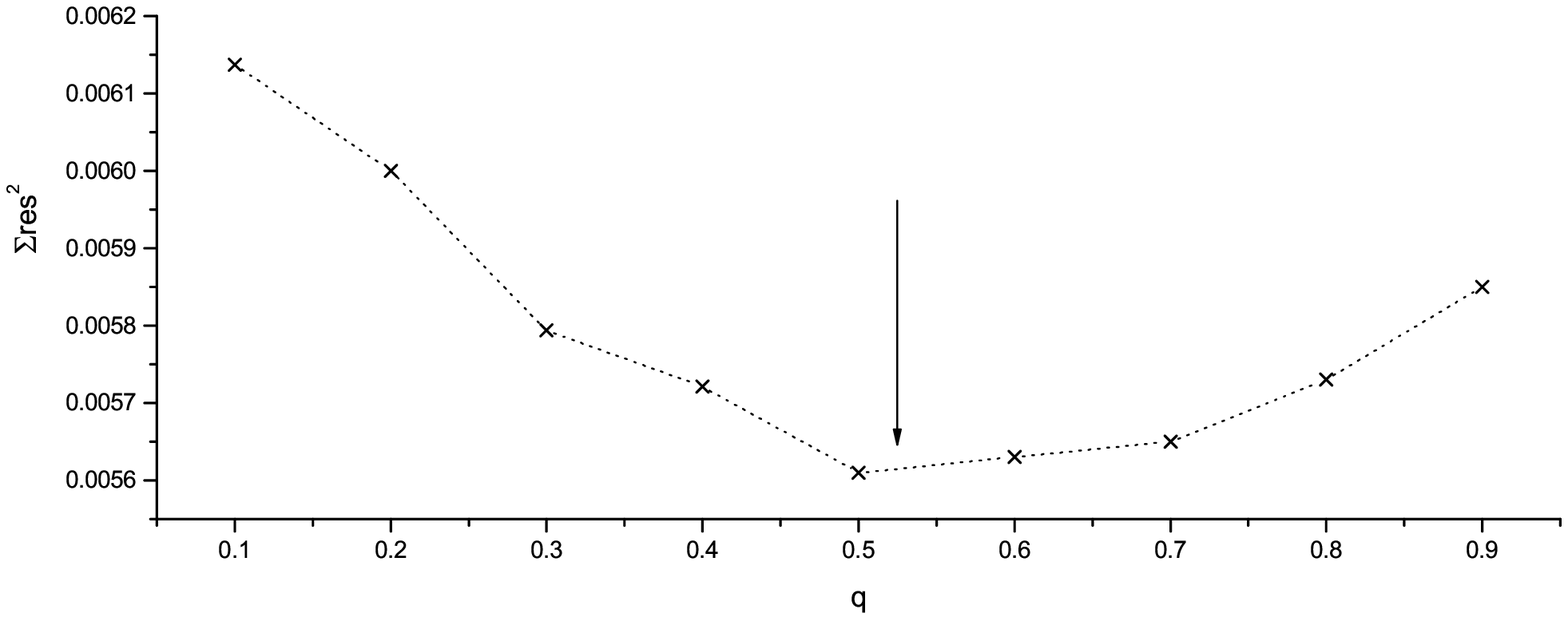}
	\caption{$q$-search plot for KIC~8504570. The arrow denotes the final adopted $q$ value after the iterations.}
	\label{fig:qs}
\end{figure}

The system was observed in long- and short-cadence modes by the $Kepler$ mission during various quarters. However, since the primary goal of this study concerns the asteroseismic analysis of the pulsating star of KIC~8504570 (i.e. pulsations modelling and mode identification), only the short-cadence data \citep[downloaded from the $KEBC$;][]{PRS11} were used for the frequency analysis. However, it should be noted that the data obtained for this system during non successive quarters of the $Kepler$ mission provide significant time gaps, something that is crucial for the frequency analysis \citep[alias effect;][]{BRE00}. Furthermore, time gaps exist also within the data of a single quarter. Therefore, the selection of data for this system was made according to their continuity and total amount in time in order to include the most possible compact data sample. More specifically, the data of Q13 and a part of Q14 were selected for analysis. In total, 150458 available points were used. These data were obtained during 106.9 consecutive days and provide 27 full LCs. The level of light contamination for this system is zero (as listed in the Mikulski Archive for Space Telescopes; MAST).
The total covering and continuous time of observations is more than three months (with negligible time gaps), which is sufficient for the study of short-period pulsations and for LC modelling. The short-cadence $Kepler$ LCs of the first 40 days of observations for KIC~8504570 are illustrated in Figure~\ref{fig:LCsandRes}. The orbital phases and the flux to magnitude conversions for this system were derived using the ephemeris ($T_{0}=2454955.78(3)$~BJD, $P_{\rm orb}=4.007705(8)$~d) and the $Kepler$ magnitude $K_{\rm p}=13.25$~mag, respectively, as listed in $KEBC$.

%%%%%PHOEBE + general LC analysis method
The LCs analyses were made with the \textsc{PHOEBE} v.0.29d software \citep{PRS05} that is based on the 2003 version of the Wilson-Devinney code \citep{WIL71, WIL79, WIL90}. The temperature ($T_{\rm eff,~1}$) of the primary component was given value as yielded from the spectral classification (see Section~\ref{sec:sp}) and was kept fixed during the analysis. On the other hand, the temperature of the secondary ($T_{\rm eff,~2}$) was adjusted. The albedos ($A$) and the gravity darkening coefficients ($g$) were assigned values according to the spectral types of the components \citep{RUC69, ZEI24, LUC67}. The (linear) limb darkening coefficients ($x$) were taken from the lists of \citet{HAM93}. The synchronicity parameters ($F$) were initially adjusted, but due to the absence of significant changes during the iterations, the system was assumed to be tidally locked (i.e. $F_1=F_2=1$) following the preliminary findings of \citet{LUR17}. The dimensionless potentials ($\Omega$), the fractional luminosity of the primary component ($L_{1}$), and the inclination of the system ($i$) were set as adjustable parameters. Since no supporting evidence for the existence of tertiary component and, additionally, since the light contamination is zero, the third light parameter ($l_3$) was not taken into account. At this point, it should be noted that the $R$ filter (Bessell photometric system--range between 550-870~nm and with a transmittance peak at 597~nm) simulates the best the spectral response of the CCD sensors of $Kepler$ (410-910~nm with a peak at $\sim588$~nm). Therefore it was used for the calculation of the filter depended parameters (i.e. $x$ and $L$) in \textsc{PHOEBE}.

%%%%%Normal points + q search
In the absence of spectroscopic mass ratio ($q$) for KIC~8504570, the $q$-search method \citep[for details see e.g.][]{LIAN12} was applied. For this, a mean LC exempted from the presence of pulsations was needed. Moreover, in this system, except the short-period pulsations, brightness variations due to magnetic activity (e.g. spots), occurring mostly in the out-of-eclipse phase parts, were also found. Therefore, the mean LC (folded into the orbital period) was calculated from two up to four successive LCs, in which there were no major brightness changes between them. It should be noted that a complete LC of  KIC~8504570 contains approximately 5500 data points. The mean LC, using averaged points per phase, contained approximately 300 normal points and the variations of both the pulsations and the spots were almost vanished. The $q$-search was applied in modes 2 (detached system), 4 (semi-detached system with the primary component filling its Roche lobe) and 5 (conventional semi-detached binary) to find feasible (`photometric') estimates of the mass ratio. The step of $q$ change during the search was 0.1 starting from $q$=0.1. The sums of the squared residuals were systematically lower for all $q$ values in mode 2, therefore, this system can be plausibly considered as a detached EB.

%%%%%%% details foir this system
According to the $q$-search method, the minimum sum of squared residuals was found for $q$=0.5 (Figure~\ref{fig:qs}). This value was initially assigned to $q$, but later on it was adjusted. This system presents remarkable brightness changes from cycle to cycle after the 10th day of observations. It was found that for 40 continuous days after the 10th day, a hot spot on the surface of the secondary component describes the individual LCs very well. The selection of the hot spot was based on the results of \citet{DAV16} regarding possible flare activity in the system and fits well to a profile of a star with temperature of 5300~K (secondary component). Between the 52nd and 75th days of observations, no spots were required for the LC model, in contrast with the time range between 76th and 104th days, for which a cool spot was adopted on the surface of the same component. The spots parameters (colatitude $Colat$, longitude $long$, radius, and temperature factor $Tf$) were adjusted in the individual LC models. Finally, for this system, one model per LC was obtained, thus, 27 models were totally derived and combined for the final average model.

%%%%% Justification for multiple models technique
The analyses of $Kepler$ LCs for EBs require special handling due to light variations caused by magnetic spots between successive LCs \citep[c.f.][]{LIA17, LIA18, LIA20}. The latter justifies our choice not to model all the available points folded into the $P_{\rm orb}$, but to model each LC separately. This method provides more realistic errors for the final model results, since its single parameter (except from those of the spots) is the average from those of the individual models, while its error is the standard deviation of them. Moreover, using this method, the brightness changes due to the spots and other proximity effects are well modeled, hence, the final LC residuals can be considered as free as possible of the binarity, something that is extremely crucial for the subsequent pulsation analysis (Section~\ref{sec:Fmdl}).

\begin{figure}[t]
	\centering
	\begin{tabular}{cc}
		\includegraphics[width=13cm]{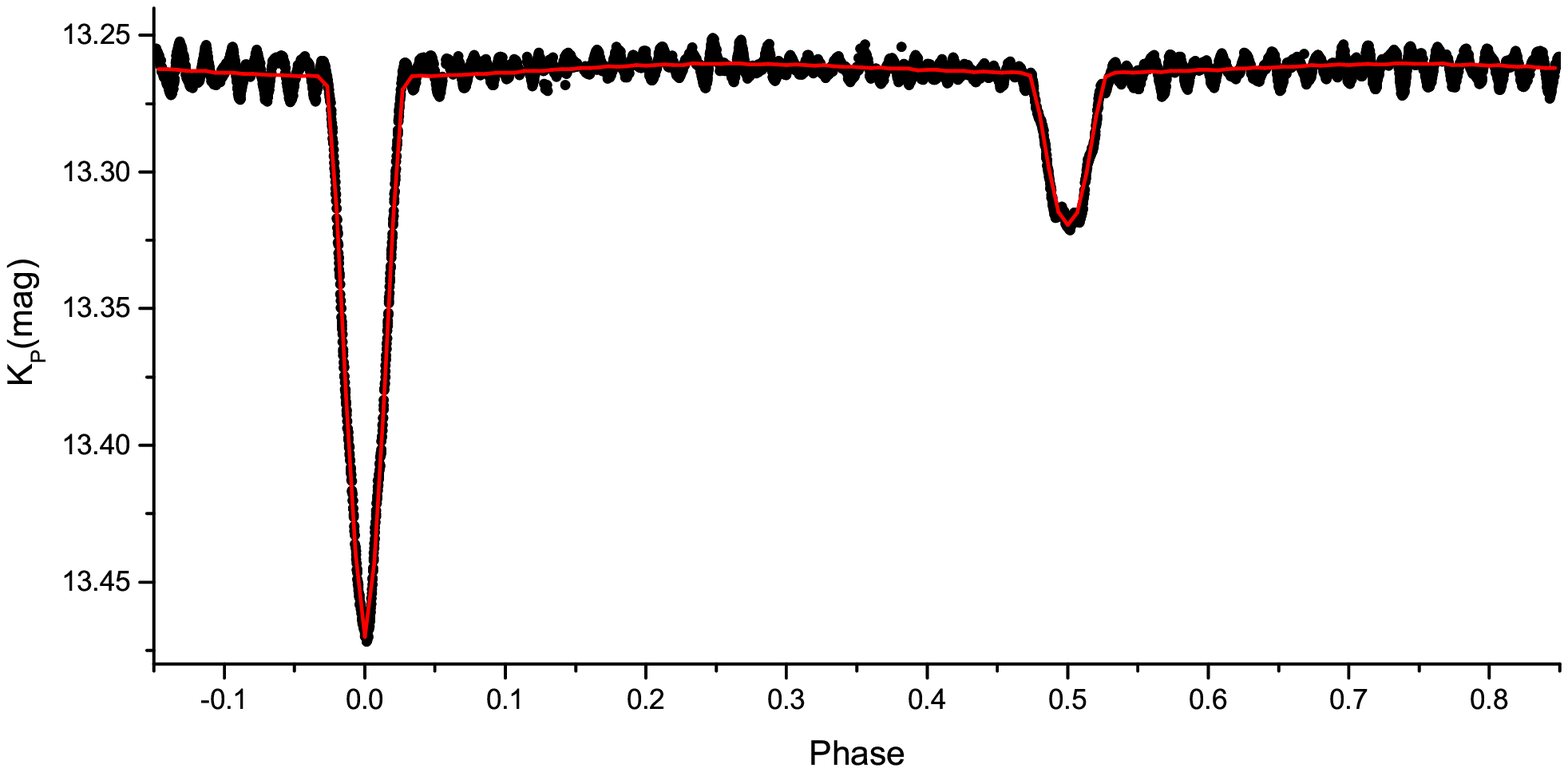}\\
		\includegraphics[width=7.5cm]{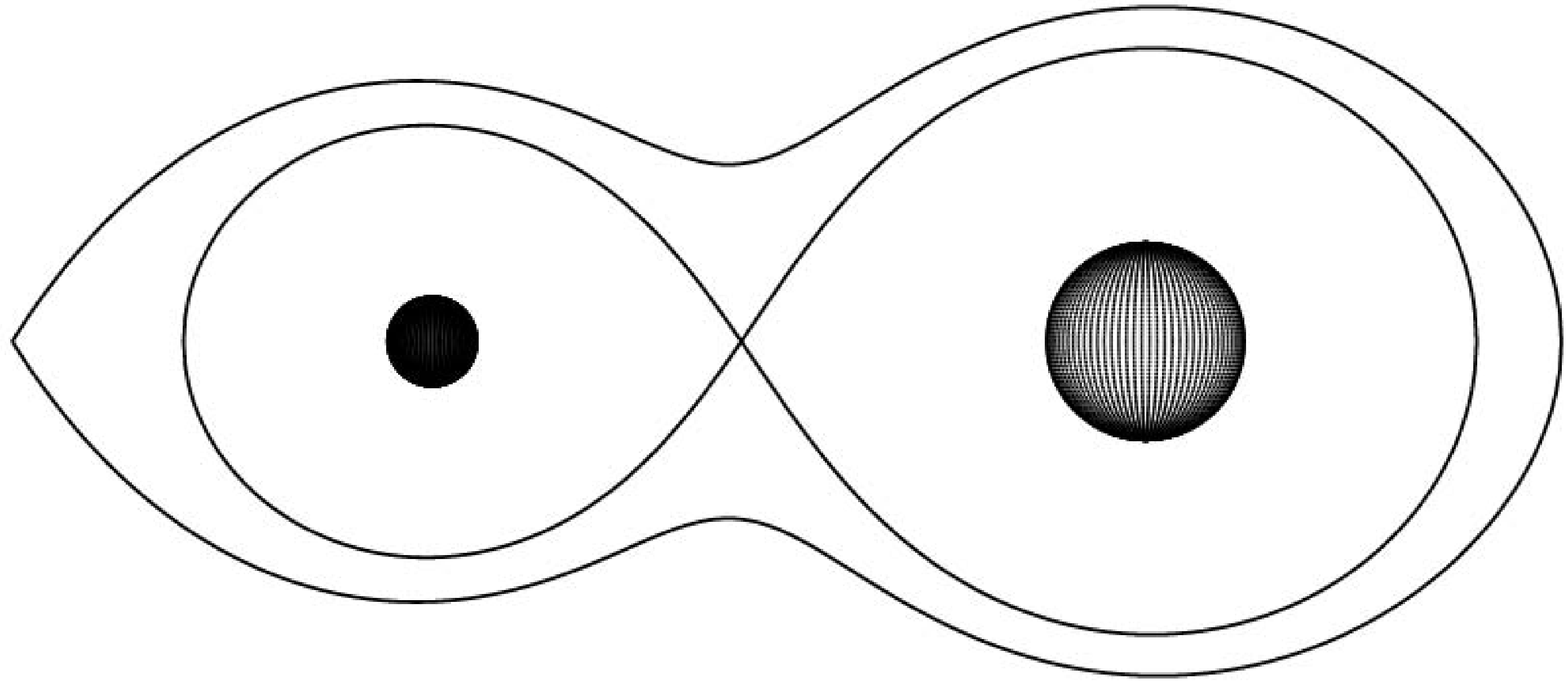}\\
	\end{tabular}
	\caption{Example of LC modelling (solid lines) over the observed $Kepler$ LCs (points) of an individual orbital cycle and the Roche geometry plot (the right star is the primary component) at orbital phase 0.75 for KIC~8504570.}
	\label{fig:LCm3D}
\end{figure}

%%%%%%%%%%%%%%%%%%%%%%%%%%%%%%%%%%%%%%%%%%%%%%%%%%%%%%%%%%%%%%%%%%%%%%%%%%%%%%%%%%%%%%%%%% Table for LC model + Absolute parameters %%%%%%%%%%%%%%%%%%%%%%%%%%%%%%%%%%%%%%%%%%%%%%%%%%%%%%%%%%%%
\begin{table}[h]
	\centering
	\caption{LCs modelling and absolute parameters for KIC~8504570. The errors are given in parentheses alongside values and correspond to the last digit(s).}
	\label{tab:LCmdlAbs}
	\begin{tabular}{ccc ccc }
\toprule											
&	\multicolumn{2}{c}{Components parameters}			&		&	\multicolumn{2}{c}{Absolute parameters}			\\
\midrule											
&	Primary	&	Secondary	&		&	Primary	&	Secondary	\\
\midrule											
$T_{\rm eff}$~(K)	&	7450(150)$^{(1)}$	&	5300(113)	&	$M~$($M_{\odot}$)	&	1.67(17)$^a$	&	0.87(9)	\\
$\Omega$	&	7.76(4)	&	9.57(4)	&	$R~$($R_{\odot}$)	&	1.97(9)	&	0.89(14)	\\
$A$	&	1$^a$	&	0.5$^a$	&	$L~$($L_{\odot}$)	&	11(1)	&	0.6(2)	\\
$g$	&	1$^a$	&	0.32$^a$	&	$\log g$~(cm~s$^{-2}$)	&	4.07(5)	&	4.48(15)	\\
$x$	&	0.455	&	0.583	&	$a$~($R_{\odot}$)	&	5.07(8)	&	9.75(4)	\\
$L$/$L_{\rm T}$	&	0.945(1)	&	0.055(2)	&	$M_{\rm bol}$~(mag)	&	2.17(6)	&	5.4(2)	\\
\cline{4-6}											
$r_{\rm pole}$	&	0.133(1)	&	0.060(1)	&		&	\multicolumn{2}{c}{System parameters}			\\
\cline{4-6}											
$r_{\rm point}$	&	0.134(1)	&	0.060(1)	&	$q$~($m_{2}$/$m_{1}$)	&		\multicolumn{2}{c}{0.52(1)}		\\
$r_{\rm side}$	&	0.133(1)	&	0.060(1)	&	$i$~($\deg$)	&		\multicolumn{2}{c}{84.6(1)}		\\
$r_{\rm back}$	&	0.133(1)	&	0.060(1)	&		&		&		\\
\bottomrule										
\end{tabular}
\\$^1$~Section~\ref{sec:sp}, $^a$assumed, $L_T = L_1+L_2$
\end{table}

%%%%%%%%%%%%%%%%%%%%%%%%%%%%%%%%%%%%%%%%%%%%%%%%%%%%%%%%%             RESULTS
The LCs modelling results for KIC~8504570 are listed in Table~\ref{tab:LCmdlAbs}. An example of LC modelling and Roche geometry representation are plotted in Figure~\ref{fig:LCm3D}. The LCs residuals after the subtraction of the individual models are illustrated below the observed LCs in Figure~\ref{fig:LCsandRes}. Moreover, the parameters of the spots for each LC (cycle) are given in Table~\ref{tab:spots} in appendix~\ref{sec:App2}. Figure~\ref{fig:spotKIC085} includes the immigration plots of the spots and their locations on the surface of the secondary component for two different dates of observations.

Although no RVs curves exist for this system, the absolute parameters of its components can be estimated making plausible assumptions. The adopted mass (1.67~$M_{\odot}$) of the primary was based on its spectral type according to the spectral type-mass correlations of \citet{COX00} for main-sequence stars. A fair mass error value of 10$\%$ was also adopted. The mass of the secondary component can be directly derived from the calculated (photometric) mass ratio. The semi-major axes $a$ can then be derived from Kepler's third law. The luminosities ($L$), the gravity acceleration ($\log g$), and the bolometric magnitudes values ($M_{\rm bol}$) were calculated using the standard definitions. The calculation of the absolute parameters was made with the software \textsc{AbsParEB} \citep{LIA15} and they are listed in Table~\ref{tab:LCmdlAbs}.

%%%%%%%%%%%%%%%%%%%%%%%%%%%%%%%%%%%%%%%%%%%%%%%%%%%%%%%%%%%%%%%%%%%%%%%%%%%%%%%%%%%% S E C T I O N 4 ----- Pulsation Frequencies analysis
%%%%%%%%%%%%%%%%%%%%%%%%%%%%%%%%%%%%%%%%%%%%%%%%%%%%%%%%%%%%%%%%%%%%%%%%%%%%%%%%%%%%
%%%%%%%%%%%%%%%%%%%%%%%%%%%%%%%%%%%%%%%%%%%%%%%%%%%%%%%%%%%%%%%%%%%%%%%%%%%%%%%%%%%%
%%%%%%%%%%%%%%%%%%%%%%%%%%%%%%%%%%%%%%%%%%%%%%%%%%%%%%%%%%%%%%%%%%%%%%%%%%%%%%%%%%%%								
\section{Pulsations modelling}
\label{sec:Fmdl}

%%%%%%%%%%%%%%%%%%%%%%%%%%%%%%%%%%%%%%%%%% General method
The search for pulsation frequencies was made with the software \textsc{PERIOD04} v.1.2 \citep{LEN05} that is based on classical Fourier analysis. Although the typical frequencies range of $\delta$~Scuti stars is 4-80~d$^{-1}$ \citep{BRE00}, the present analysis included also the regime 0-4~d$^{-1}$. This selection was based on the fact that it has been noticed \citep[e.g.][]{LIA17, LIA20} that these stars may also exhibit longer-period oscillations due either to tidal effects, that are connected to their $P_{\rm orb}$, or even to intrinsic hybrid behaviour of $\gamma$~Doradus-$\delta$~Scuti type. Therefore, the present pulsation analysis was made in the range 0-80~d$^{-1}$ on the LCs residuals of the system (Figure~\ref{fig:LCsandRes}). Moreover, since the eclipses affect the amplitudes of the pulsations (i.e. variations of the total light) and in order to keep the data sample homogeneous, only the out-of-eclipse data were used. The ranges of orbital phases ($\Phi_{\rm orb}$) of the excluded data were 0.97-0.03 and 0.47-0.53. For the signal-to-noise ratio (S/N) calculation of the frequencies, the method for the background noise estimation, as described in detail in \citet{LIA17}, was applied. Particularly, the background noise of the data set was calculated as 7.51~$\upmu$mag in regimes with absence of frequencies, with a spacing of 2~d$^{-1}$ and a box size of 2. A 4$\upsigma$ limit \citep[S/N$=4$;][]{LEN05} regarding the reliability of the detected frequencies was adopted (0.03~mmag). Hence, after the first frequency computation the residuals were subsequently pre-whitened for the next one until the detected frequency had S/N$\sim$4. The Nyquist frequency and the frequency resolution according to the Rayleigh-Criterion (i.e. 1/$T$, where $T$ is the observations time range in days; c.f. \citet{AER10}, \citet{SCH03}) for the present data set were 239.5~d$^{-1}$ and 0.009~d$^{-1}$, respectively. According to the present spectroscopic and LCs modelling results (Sections~\ref{sec:sp} and \ref{sec:LCmdl}), only the primary of KIC~8504570 simulates adequately the properties of $\delta$~Scuti type stars (i.e. mass and temperature), hence, it can be plausibly concluded that this star is the pulsator of this system.

%%%%%%%%%%%%%%%%%%%%%%%%%%%%%%%%%%%%%%%%%frequency modes
After the frequency search, the pulsation constant for each independent frequency ($f$) was calculated based on the relation of \citet{BRE00}:
\begin{equation}
\log Q = -\log f + 0.5 \log g + 0.1M_{\rm bol} + \log T_{\rm eff}- 6.456.
\label{eq:Q}
\end{equation}
Moreover, the following pulsation constant-density relation was used for the calculation of the density of the pulsators:
\begin{equation}
Q = f_{\rm dom}^{-1}\sqrt{\rho_{\rm pul}/\rho_{\odot}},
\label{eq:rho}
\end{equation}
where $f_{\rm dom}$ is the frequency of the dominant pulsation mode (i.e. that with the largest amplitude). At this point it should be noted that the $f_{\rm dom}$ of the multiperiodic $\delta$~Scuti stars varies over time. Therefore, for a more realistic estimation of the density of this pulsator, the average value of $Q$ of the independent frequencies was used.

The identification of the oscillating modes (i.e. $l$-degrees and type) employed the theoretical MAD models for $\delta$~Scuti stars \citep{MON07} in the \textsc{FAMIAS} software v.1.01 \citep{ZIM08}. The $l$-degrees from the closest MAD models (i.e. $f$, $\log g$, $M$, and $T_{\rm eff}$) to the detected independent frequencies were adopted as the most possible pulsation modes. Moreover, the ratio $P_{\rm pul}$/$P_{\rm orb}$ of all independent frequencies was calculated in order to check if it is less than 0.07, which is the upper value, according to \citet{ZHA13}, for the discrimination of $p$-type modes.

\begin{figure}[h!]
	\centering
	\begin{tabular}{c}
	\includegraphics[width=13cm]{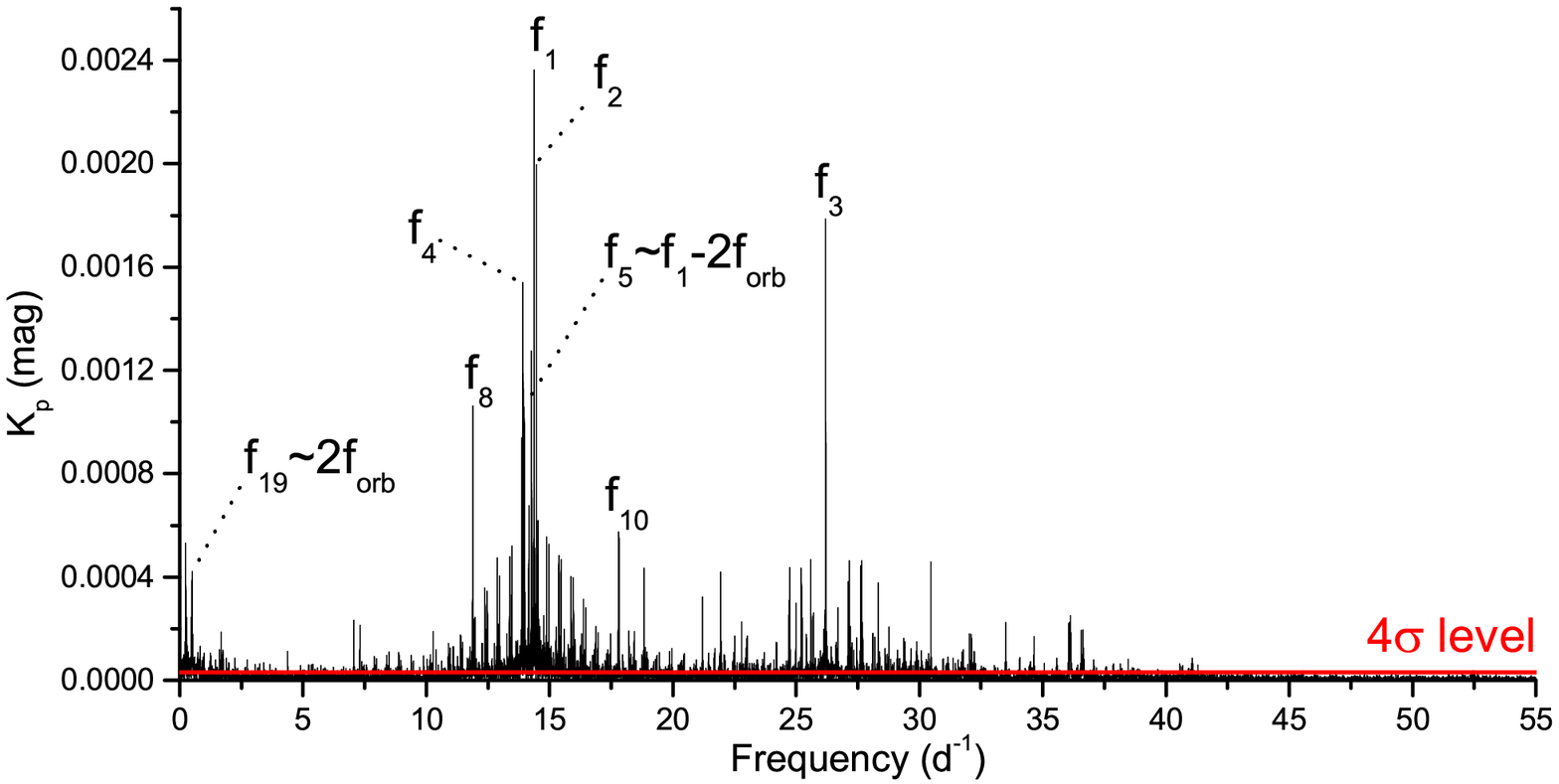}\\
	\includegraphics[width=13cm]{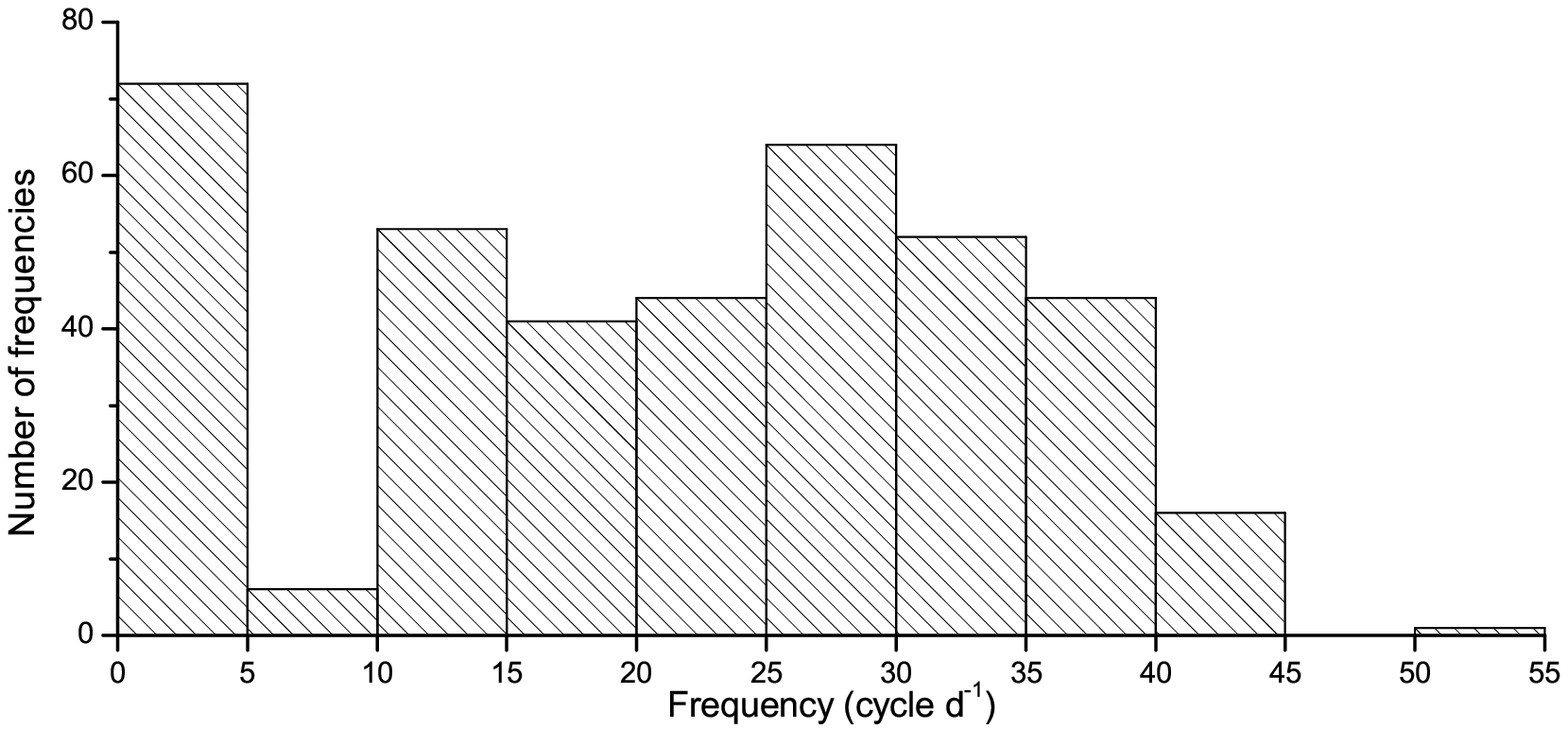}\\
	\end{tabular}
	\caption{Periodogram (top) and frequencies distribution (bottom) for KIC~8504570. The independent frequencies, the strong frequencies that are connected to the $P_{\rm orb}$, and the significance level are also indicated.}
	\label{fig:FS}
	\vspace{0.5cm}
%\end{figure}
%\begin{figure}
	\centering
	\begin{tabular}{c}
		\includegraphics[width=\columnwidth]{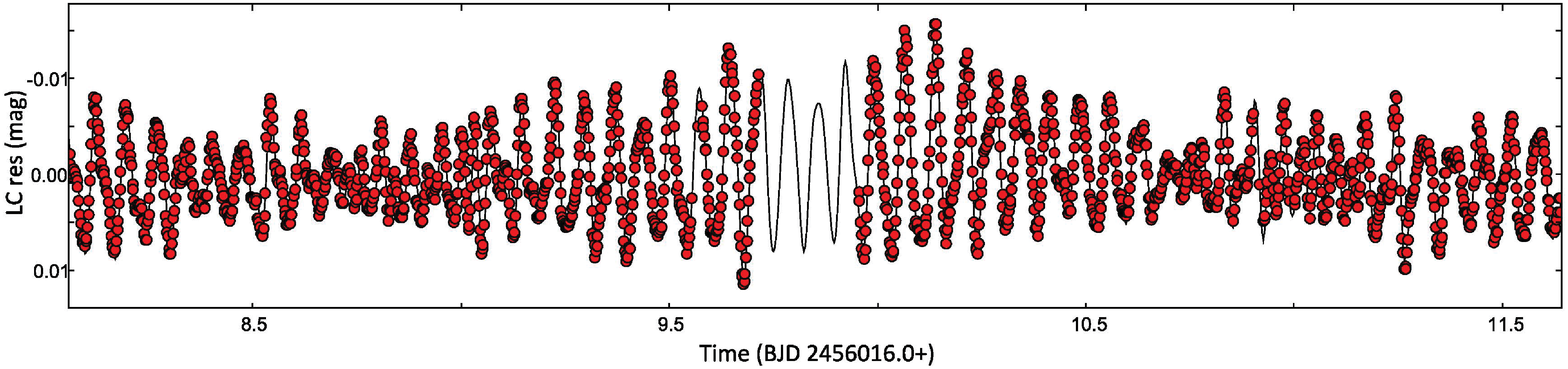}\\
	\end{tabular}
	\caption{Representative example of Fourier fitting (solid line) on various data points for KIC~8504570.}
	\label{fig:FF}
\end{figure}

%%%%%%%%%%%%%%%%%%%%%%%%%%%%%%%%%%%%%%%%%Results figures and Tables
Table~\ref{tab:IndF} includes the pulsation modelling results regarding the independent frequencies for KIC~8504570 as well as their respective mode identification. Particularly, this table lists: The frequency value $f_{\rm i}$, the amplitude $A$, the phase $\Phi$, the S/N, the $Q$, the $P_{\rm pul}$/$P_{\rm orb}$, the $l$-degrees, and the mode of each detected independent frequency. The rest detected frequencies (i.e. dependent/combination frequencies) are given in appendix~\ref{sec:App1} (Table~\ref{tab:DepFreqKIC085}). Figure~\ref{fig:FS} shows the periodogram of the pulsating star of KIC~8504570 and the distribution of its oscillation frequencies. Representative Fourier fittings on the LCs residuals are plotted in Figure~\ref{fig:FF}.

%%%%%%%%%%%%%%%%%%%%%%%%%%%%%%%%%%%%%%%%%%%%%%%%%%%%%%%%%%%%%%%%%%%%%%%%%%%%%%%%%%%%%%%%% Table for Indepedent Freqs %%%%%%%%%%%%%%%%%%%%%%%%%%%%%%%%%%%%%%%%%%%%%%%%%%%%%%%%%%%%
\begin{table}
	\centering
	\caption{Frequency search results and mode identification for the independent frequencies of the pulsating component of KIC~8504570. The errors are given in parentheses alongside values and correspond to the last digit(s).}
	\label{tab:IndF}
	\begin{tabular}{l cccc cccc}
		\toprule																	
		$i$	&	  $f_{\rm i}$	&	$A$	&	  $\Phi$	&	S/N	  & 	$Q$	&	         $P_{\rm pul}$/$P_{\rm orb}^{\rm 1}$	&	$l$-degree	&	Pulsation mode	\\
		&	     (d$^{-1}$)	&	(mmag)	&	($^\circ$)	&		  & 	(d)     	&		&				\\
		\midrule															
		1	&	14.37417(1)	&	2.550(5)	&	1.8(1)	&	339.3	&	0.032(2)	&	0.017	&	3 or 1	&	NR $p$	\\
		2	&	14.46668(1)	&	2.215(5)	&	59.2(1)	&	294.7	&	0.032(2)	&	0.017	&	1 or 3	&	NR $p$	\\
		3	&	26.19524(1)	&	1.779(5)	&	32.1(1)	&	236.7	&	0.018(1)	&	0.010	&	2	&	NR $p$	\\
		4	&	13.91847(2)	&	1.560(5)	&	236.5(2)	&	207.6	&	0.033(2)	&	0.018	&	2 or 3	&	NR $p$	\\
		8	&	11.89034(2)	&	1.070(5)	&	57.1(2)	&	142.4	&	0.039(2)	&	0.021	&	3	&	NR $p$	\\
		10	&	17.80685(4)	&	0.565(5)	&	264.1(5)	&	75.2	&	0.026(2)	&	0.014	&	1	&	NR $p$	\\
		\bottomrule																						
	\end{tabular}
	\\$^1$Error values are of 10$^{-7}$-10$^{-8}$ order of magnitude. NR $p$ non-radial pressure mode
\end{table}

The pulsator of KIC~8504570 oscillates in a total of 393 frequencies. Six of them are independent and were detected in the regime 11.8-26.2~d$^{-1}$. From the rest 387 depended frequencies, the 309 are spread almost uniformly in the range 10-43.4~d$^{-1}$, 72 have values less than 4.4~d$^{-1}$, five were found between 5.5-9~d$^{-1}$, while only one, namely $f_{282}$, exceeds the 50~d$^{-1}$. As can be seen in Figure~\ref{fig:FS}, one main concentration of frequencies is between 12-17~d$^{-1}$, while a slightly more spread one is between 23-30~d$^{-1}$. The results based on MAD models show that all oscillations are probably non-radial pressure modes. Although the ratio $f_4$/$f_{10}$ has value $\sim0.78$, thus, $f_4$ was not identified as radial mode by the MAD models. Finally, a value of $\rho_{\rm pul}$=0.215(4)~$\rho_{\odot}$ was derived.

\newpage
\section{Summary, discussion and conclusions}
\label{sec:Dis}

%%%%%%%%%%%%%%%%%%%%%%%%%%%%%%%%%%%%%%%%%%%%%%%%%%%Summary
In the present work, detailed LCs and pulsations modellings for KIC~8504570, a neglected $Kepler$ detached EB with an oscillating component, are presented. The spectral classification of its primary component, based on our spectroscopic observations, provided the means for accurate LCs analyses, as well as for the estimation of the absolute parameters and evolutionary stages of both the components of the EB. The primary component was also identified as a $\delta$~Scuti star and its pulsational characteristics (pulsation frequencies model and mode identification) were accurately determined.

%%%%%%%%%%%%%%%%%%%%%%%%%%%%%%%%%%%%%%%%%%%%%%%%%%%KIC 085
The primary component of KIC~8504570 was classified as an A9 type star and pulsates in six independent frequencies in the regime 11.89-26.2~d$^{-1}$ with the dominant one at 14.37~d$^{-1}$. These frequencies were identified as non-radial (pressure) modes according to the MAD models. Moreover, this star oscillates in another 387 combination frequencies. During the LC modelling, initially a hot and subsequently a cool spot on the surface of the secondary component were used to overcome the brightness asymmetries in the quadratures. This selection can be justified from the fact that this EB was listed as a possible flare system \citep{DAV16}.

\begin{figure}[h]
	\centering
	\includegraphics[width=13cm]{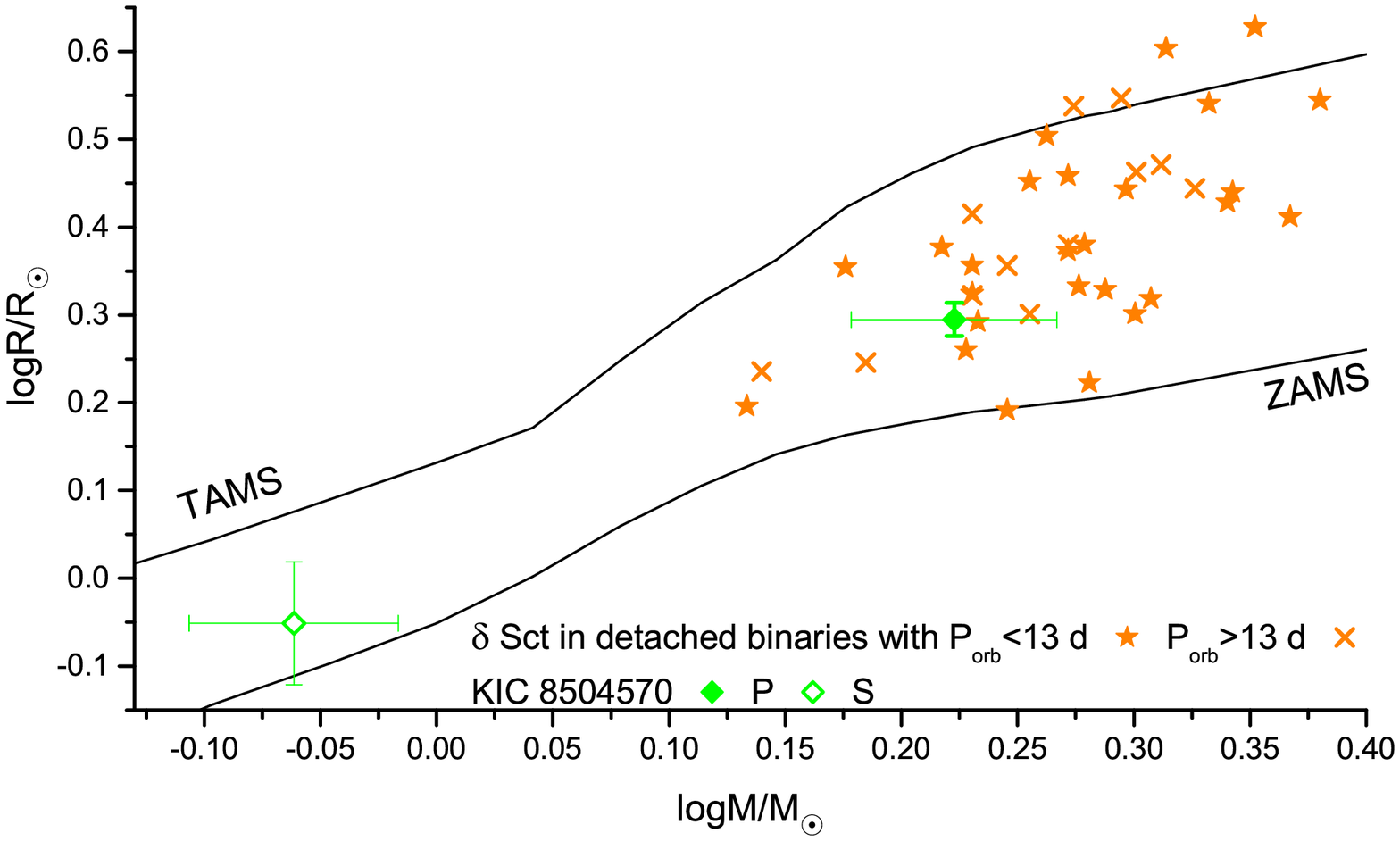}
	\caption{Locations of the primary (filled symbol) and secondary (empty symbol) components of KIC~8504570 (diamonds) within the mass-radius diagram. The stars and the `x' symbols denote the $\delta$~Scuti components of other detached systems with $P_{\rm orb}$ shorter and longer than 13~d, respectively (taken from \citet{LIAN17} and \citet{LIA20}). The black solid lines represent the main-sequence edges.}
	\label{fig:MR}
	\end{figure}

\begin{figure}
		\centering
	\includegraphics[width=13cm]{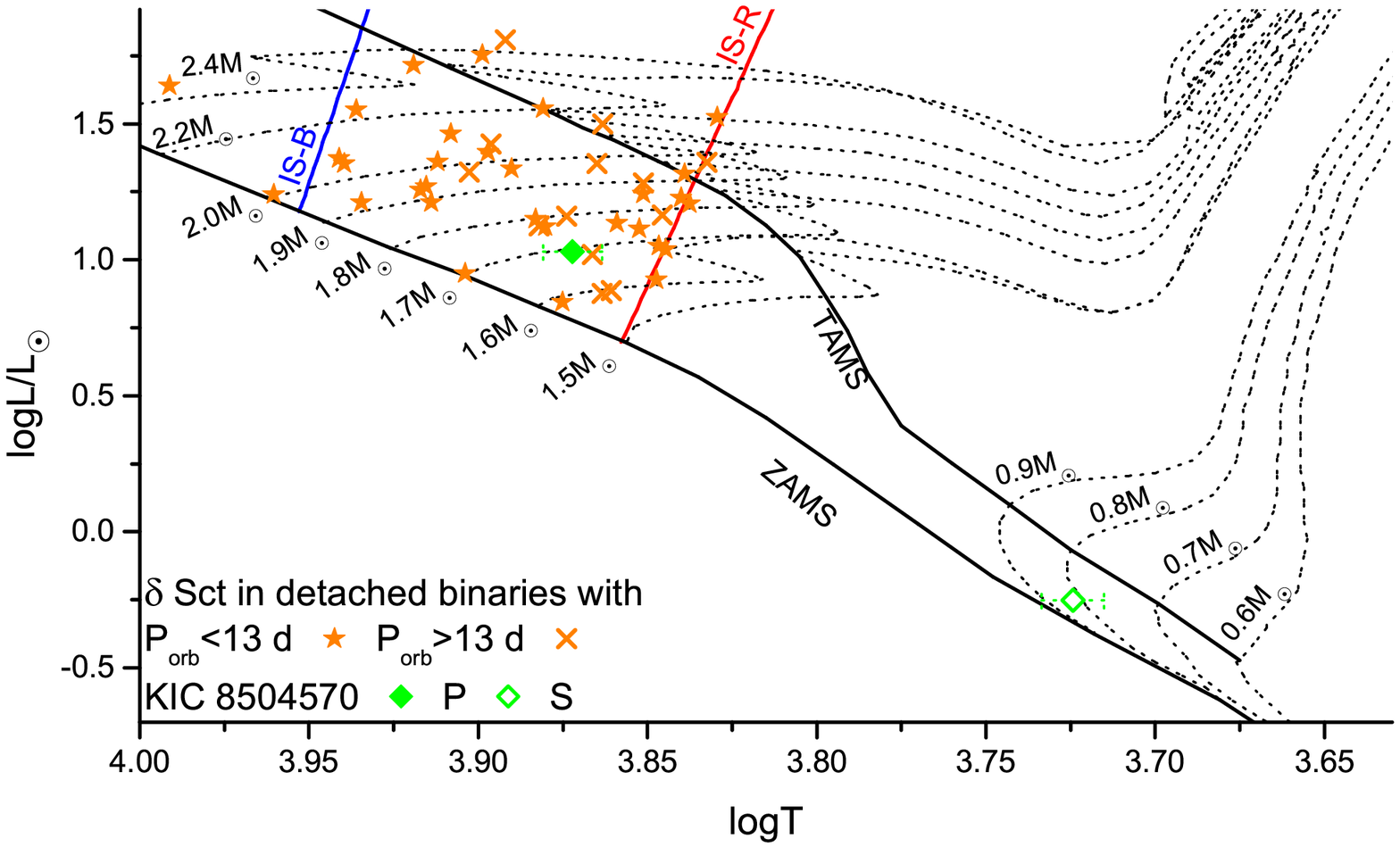}
	\caption{Locations of the components of KIC~8504570 within the $HR$ diagram. Symbols and black solid lines have the same meaning as in Figure~\ref{fig:MR}. Dotted lines denote the evolutionary tracks for stars with masses between 0.6-0.9~$M_{\odot}$ and 1.5-2.4~$M_{\odot}$ (taken from \citet{GIR00}) and the coloured solid lines (B=Blue, R=Red) the boundaries of the instability strip (IS; taken from \citet{SOY06b}).}
	\label{fig:HR}
\end{figure}

%%%%%%%%%%%%%%%%%% H-R M-R
For the estimation of the evolutionary stages of the components of KIC~8504570, the locations of its members on the mass-radius ($M-R$) and Hertzsprung-Russell ($HR$) diagrams are illustrated in Figures~\ref{fig:MR} and \ref{fig:HR}, respectively. Both components are located inside the main-sequence and follow very well the theoretical evolutionary tracks of \citet{GIR00} (see Figure~\ref{fig:HR}) according to their derived masses and the corresponding error ranges (see Table~\ref{tab:LCmdlAbs}). Therefore, it seems that they have been evolving without any significant interactions so far. In terms of evolution, the $\delta$~Scuti component of KIC~8504570 has similar absolute properties with other $\delta$~Scuti stars in detached binary systems. It is among the eight less massive and less luminous stars of this sample and it is located closer to red edge of the classical instability strip.


\begin{figure}[h!]
	\centering
	\includegraphics[width=13cm]{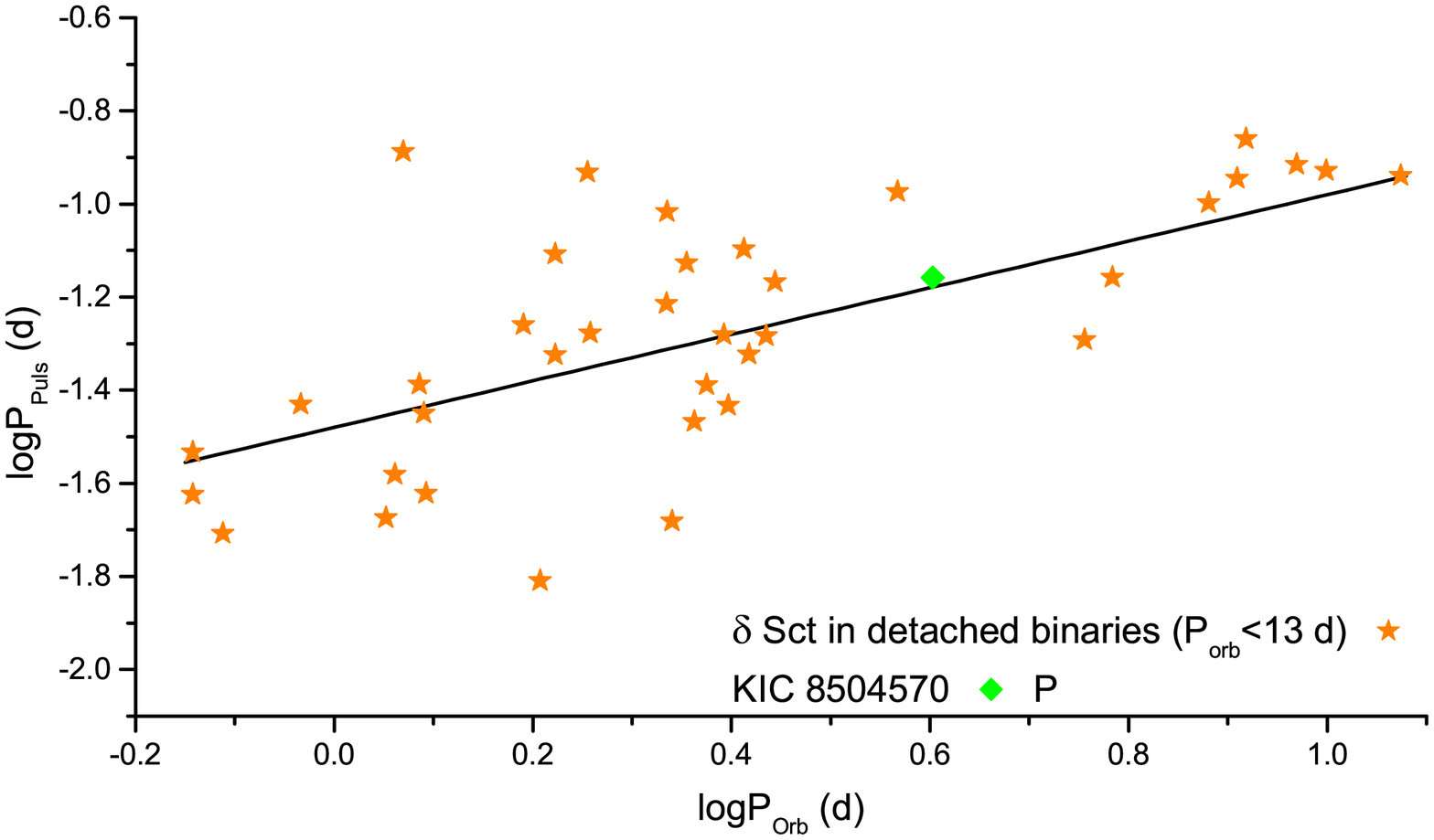}
	\caption{Location of the oscillating (primary) component of KIC~8504570 among other $\delta$~Scuti stars-members of detached systems with $P_{\rm orb}<13$~d within the $P_{\rm orb}-P_{\rm pul}$ diagram. Symbols have the same meaning as in Figure~\ref{fig:MR}, while the solid line denotes the empirical linear fit of \citet{LIA20}.}
	\label{fig:PP}
\end{figure}

\begin{figure}[h!]
	\centering
	\includegraphics[width=13cm]{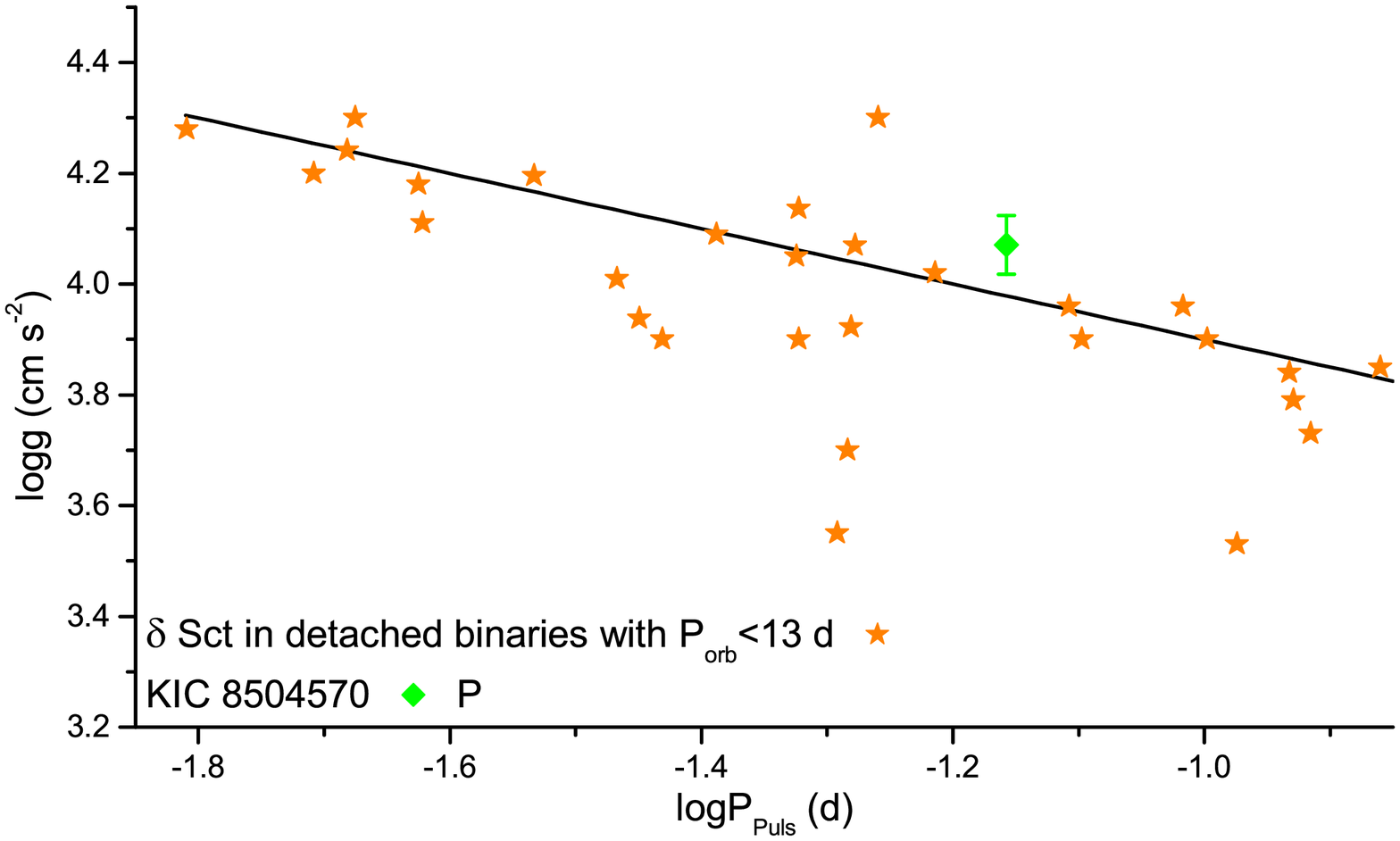}
	\caption{Location of the $\delta$~Scuti (primary) star of KIC~8504570 within the $\log g-P_{\rm pul}$ diagram. Symbols and lines have the same meaning as in Figure~\ref{fig:PP}.}
	\label{fig:gP}
\end{figure}

%%%%%%%%%%%%%%%% P-P and g-P diagrams
In order to check the accordance of the pulsational properties of the $\delta$~Scuti star of KIC~8504570 with others that belong in similar systems, it is placed on the $P_{\rm pul}-P_{\rm orb}$ and $\log g - P_{\rm pul}$ diagrams (Figures~\ref{fig:PP} and \ref{fig:gP}, respectively) along with the well established empirical relations of \citet{LIA20} for $\delta$~Scuti stars in detached binaries with $P_{\rm orb}<13$~d. The studied star in these plots follows very well both the distributions of the sample stars as well as the empirical relations.


%%%%%%%%%%%%%%%%%%%%%%%%%%%%%%%%%%%%%%%%%%%%%%%%%%%Distances
Using the current dominant oscillation frequency of the pulsator and the pulsation period-luminosity relation for $\delta$~Scuti stars of \citet{ZIA19}:
\begin{equation}
M_{\rm V} = -2.94(6) \log P_{\rm pul}-1.34(6),
\label{EQ:PL}
\end{equation}
it is feasible to calculate its absolute magnitude ($M_{\rm V}$=2.06(13)~mag). Hence, using the apparent magnitude ($m_{\rm V}$) and the distance modulus, its distance can be calculated. The $m_{\rm V}$ of KIC~8504570 is 13.28~mag according to the NOMAD-1 catalog \citep{ZAC04},the extinction in $V$ band is $A_{\rm V}=0.336$~mag \citep{BER18}, thus, its distance is determined as 1502$^{+93}_{-87}$~pc. This value is in very good agreement with the value 1488$\pm 41$~pc as derived by \citet{BER18} and \citet{BAI18} and in slight disagreement with the value 1305~pc of \citet{QUE20}. The latter discrepancy is attributed to the different extinction value ($A_{\rm V}=0.474$~mag) used by \citet{QUE20}. It should be noted that the aforementioned $M_{\rm V}$ is in very good agreement with the $M_{\rm bol, 1}$=2.17(6)~mag, which was calculated based on the LC modeling (Table~\ref{tab:LCmdlAbs}).

%%%%%%%%%%%%%%%%%%%%%%%%%%%future
For the future, radial velocity measurements are welcome to validate the present results for the LC model, although the $\sim95\%$ light domination of the primary component makes the acquisition of the radial velocities of the secondary an extremely difficult task. At best, we anticipate that only the radial velocities of the primary can be measured, which will only constrain the mass of the primary component, hence the mass ratio of the system. However, these potential future measurements cannot significantly change the present pulsations models, especially the results for the dominant and the independent frequencies, which were the main goals of the present study. The asteroseismic modelling of other similar systems, especially of those observed by satellite missions, is highly encouraged and recommended because the sample of $\delta$~Scuti stars in binary systems is still small and we lack of enough information. Moreover, systems with $P_{\rm orb}$ between 10-20~d should be prioritized for detailed analysis in order to check the reasons for the existence of the boundary of $P_{\rm orb} \sim 13$~d.

%%%%%%%%%%%%%%%%%%%%%%%%%%%%%%%%%%%%%%%%%%
\vspace{6pt}

%%%%%%%%%%%%%%%%%%%%%%%%%%%%%%%%%%%%%%%%%%
%% optional
%\supplementary{The following are available online at \linksupplementary{s1}, Figure S1: title, Table S1: title, Video S1: title.}


%%%%%%%%%%%%%%%%%%%%%%%%%%%%%%%%%%%%%%%%%%
%\authorcontributions{For research articles with several authors, a short paragraph specifying their individual contributions must be provided. The following statements should be used ``Conceptualization, X.X. and Y.Y.; methodology, X.X.; software, X.X.; validation, X.X., Y.Y. and Z.Z.; formal analysis, X.X.; investigation, X.X.; resources, X.X.; data curation, X.X.; writing--original draft preparation, X.X.; writing--review and editing, X.X.; visualization, X.X.; supervision, X.X.; project administration, X.X.; funding acquisition, Y.Y. All authors have read and agreed to the published version of the manuscript.'', please turn to the  \href{http://img.mdpi.org/data/contributor-role-instruction.pdf}{CRediT taxonomy} for the term explanation. Authorship must be limited to those who have contributed substantially to the work reported.}

%%%%%%%%%%%%%%%%%%%%%%%%%%%%%%%%%%%%%%%%%%
\funding{This research was funded by the European Space Agency (ESA) under the Near Earth object Lunar Impacts and Optical TrAnsients (NELIOTA) programme, contract no. 4000112943}

%%%%%%%%%%%%%%%%%%%%%%%%%%%%%%%%%%%%%%%%%%
\acknowledgments{The authors wish to thank Mrs Maria Pizga for proofreading the text and the three anonymous reviewers for their fruitful comments. The `Aristarchos' telescope is operated on Helmos Observatory by the Institute for Astronomy, Astrophysics, Space Applications and Remote Sensing of the National Observatory of Athens. This research has made use of NASA's Astrophysics Data System Bibliographic Services, the SIMBAD, the Mikulski Archive for Space Telescopes (MAST), and the $Kepler$ Eclipsing Binary Catalog data bases.}

%%%%%%%%%%%%%%%%%%%%%%%%%%%%%%%%%%%%%%%%%%
%\conflictsofinterest{Declare conflicts of interest or state ``The authors declare no conflict of interest.'' Authors must identify and declare any personal circumstances or interest that may be perceived as inappropriately influencing the representation or interpretation of reported research results. Any role of the funders in the design of the study; in the collection, analyses or interpretation of data; in the writing of the manuscript, or in the decision to publish the results must be declared in this section. If there is no role, please state ``The funders had no role in the design of the study; in the collection, analyses, or interpretation of data; in the writing of the manuscript, or in the decision to publish the results''.}

%%%%%%%%%%%%%%%%%%%%%%%%%%%%%%%%%%%%%%%%%%
%% optional
\abbreviations{The following abbreviations are used in this manuscript:\\

\noindent
\begin{tabular}{@{}ll}
BJD & Barycentric Julian date\\
EB & Eclipsing binary system\\
ETV & Eclipse timing variations\\
KEBC & Kepler Eclipsing binary catalog\\
KIC & Kepler Input catalog\\
LC & Light curve\\
\end{tabular}}

%%%%%%%%%%%%%%%%%%%%%%%%%%%%%%%%%%%%%%%%%%
%% optional
\appendixtitles{yes} % Leave argument "no" if all appendix headings stay EMPTY (then no dot is printed after "Appendix A"). If the appendix sections contain a heading then change the argument to "yes".
\appendix
\section{Spots migration}
\label{sec:App2}
\begin{figure}[h]
	\centering
	\begin{tabular}{lr}
		\multicolumn{2}{c}{\includegraphics[width=9cm]{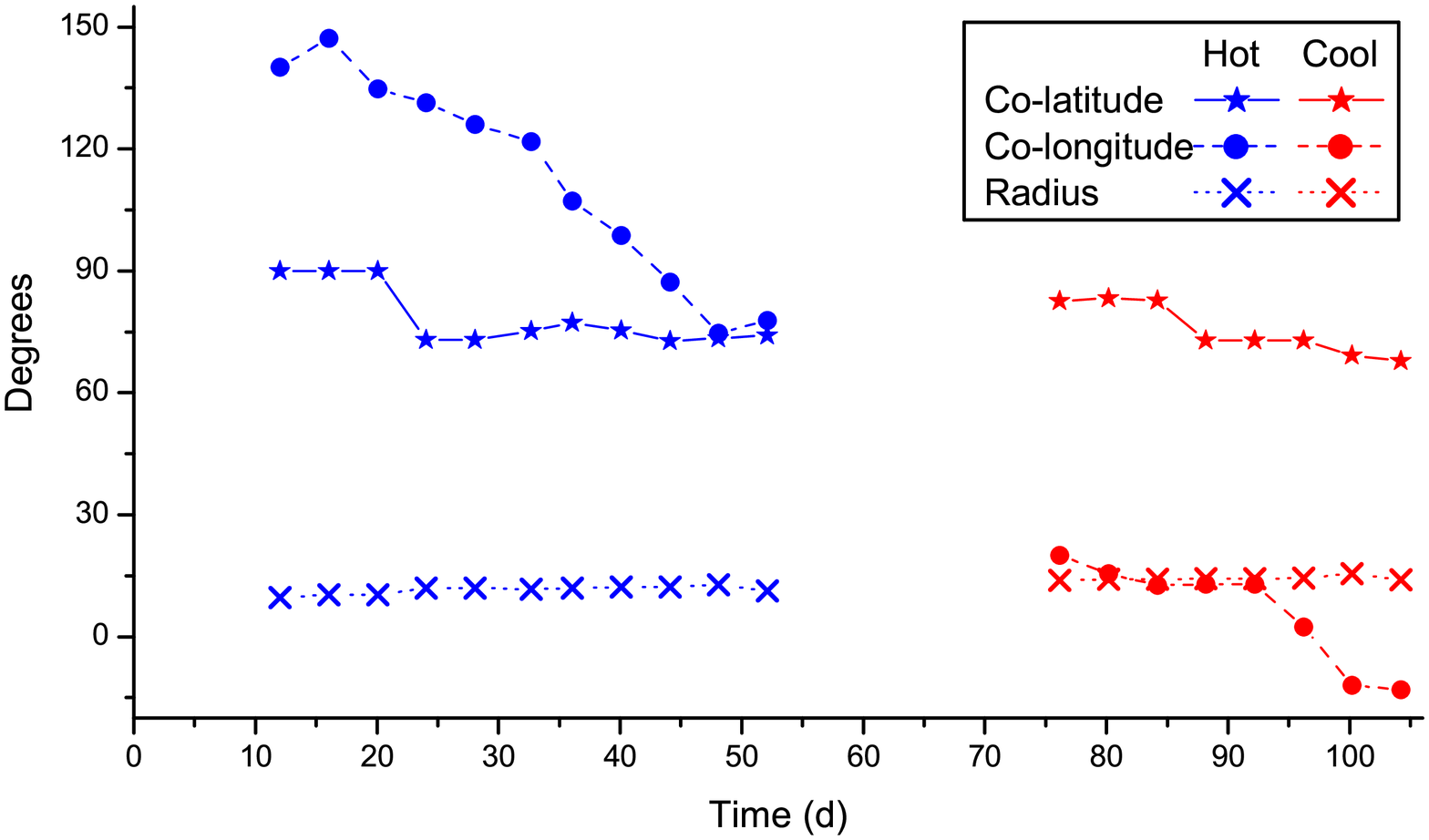}}\\
		\includegraphics[width=3cm]{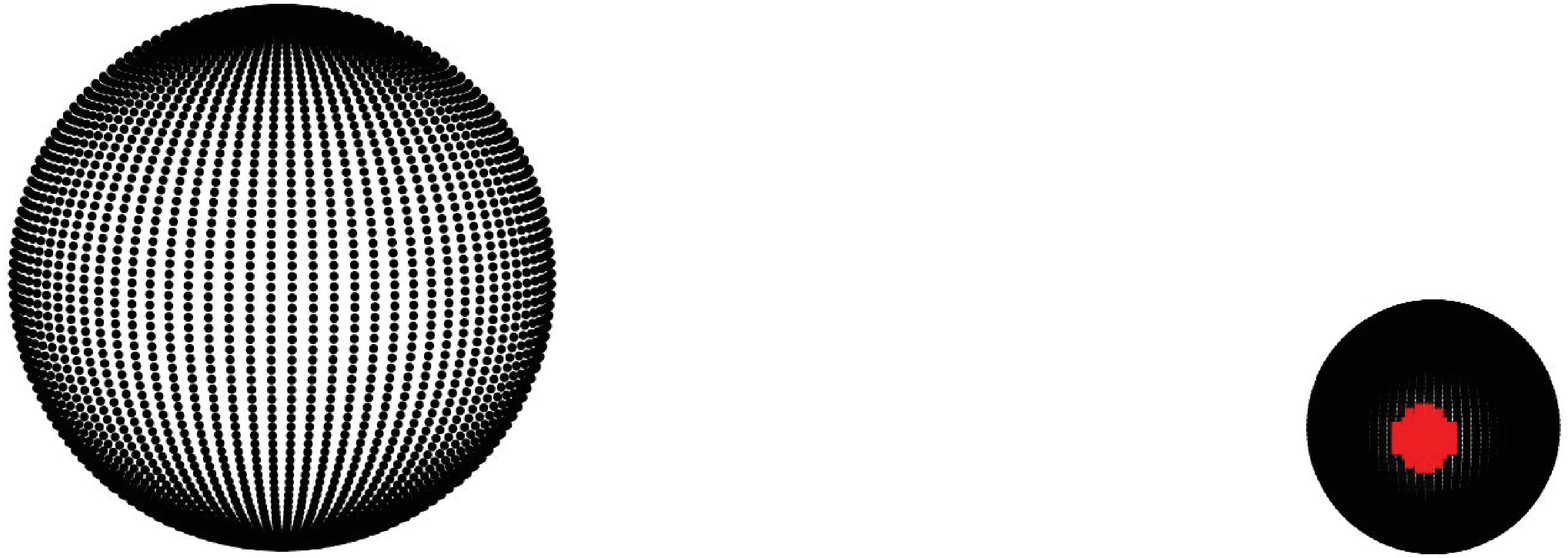}&\includegraphics[width=3cm]{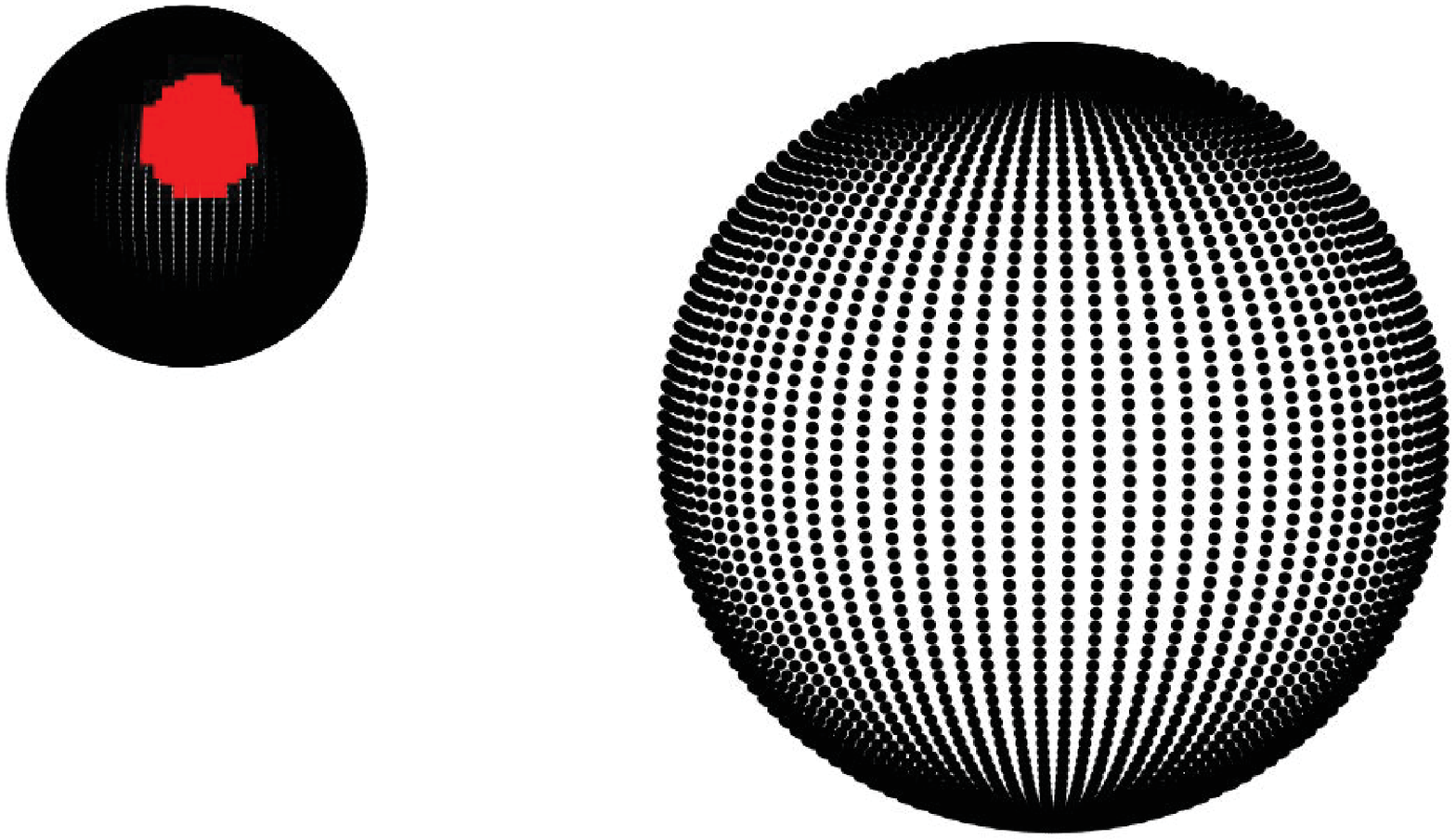}\\
	\end{tabular}
	\caption{Top panel: Spot migration diagram for KIC~8504570. Lower left panel shows the hot spot on the surface of the secondary at orbital phase 0.10 during the 12th day, while lower right panel illustrates the cool spot on the same component at an orbital phase of 0.55 during the 104th day of observations.}
	\label{fig:spotKIC085}
\end{figure}

\unskip
This appendix includes information for the variation of the spots locations in time, that were assumed to be located on the surface of the secondary of the system (see also Section~\ref{sec:LCmdl}). It should be noted, that the present solution is just a suggestion for describing the LC asymmetries in the quadratures, and more solutions (e.g. more spots with different sizes and temperatures) may result in the same LC behaviour (degeneracy of solutions). The average BJD values of the points included in the models, from which the respective parameters (colatitude $Colat.$, longitude $long.$, radius, and temperature factor $Tf$) were calculated, were set as corresponding timings for each cycle in Table~\ref{tab:spots}. The upper part of Figure~\ref{fig:spotKIC085} shows the changes of the parameters of all spots over time for the system, while the lower parts show the spot(s) on the secondary's surface during different days of observations.

%%%%%%%%%%%%%%%%%%%%%%%%%%%%%%%%%%%%%%%%%%Spot migration
\begin{table}[h]
	\centering
	\caption{Spot parameters for KIC~8504570.}
	\label{tab:spots}
\scalebox{0.93}{
	\begin{tabular}{c cc cc c cc cc}
		\toprule	
		Time	&Colat.	&	long.	&	Radius	&	Tf ($\frac{T_{\rm spot}}{T_{\rm eff}}$)	&	Time	&Colat.	&long.	&	Radius	&	Tf ($\frac{T_{\rm spot}}{T_{\rm eff}}$)	\\
		(BJD 2456016.0+)	&	($^\circ$)	&	($^\circ$)	&	($^\circ$)	&		&	(BJD 2456016.0+)	&	($^\circ$)	&	($^\circ$)	&	($^\circ$)	&		\\
		\midrule
				12.04	&	90(6)	&	140(6)	&	10(2)	&	1.18(5)	&	52.12	&	74(6)	&	78(6)	&	11(2)	&	1.09(4)	\\
		16.05	&	90(5)	&	147(5)	&	10(1)	&	1.25(2)	&	76.16	&	83(2)	&	20(4)	&	14(1)	&	0.81(6)	\\
		20.05	&	90(6)	&	135(6)	&	10(2)	&	1.25(4)	&	80.17	&	83(2)	&	16(1)	&	14(1)	&	0.72(12)	\\
		24.06	&	73(5)	&	131(5)	&	12(1)	&	1.24(2)	&	84.18	&	83(1)	&	13(2)	&	14(1)	&	0.69(16)	\\
		28.07	&	73(5)	&	126(5)	&	12(1)	&	1.24(3)	&	88.19	&	73(2)	&	13(2)	&	14(1)	&	0.65(11)	\\
		32.69	&	75(3)	&	122(2)	&	12(1)	&	1.27(4)	&	92.19	&	73(1)	&	13(2)	&	14(1)	&	0.64(11)	\\
		36.08	&	77(1)	&	107(3)	&	12(1)	&	1.25(4)	&	96.20	&	73(4)	&	2(4)	&	14(4)	&	0.57(8)	\\
		40.09	&	75(2)	&	99(3)	&	12(1)	&	1.22(5)	&	100.21	&	69(3)	&	-12(2)	&	15(3)	&	0.77(5)	\\
		44.10	&	73(9)	&	87(5)	&	12(1)	&	1.17(4)	&	104.22	&	68(5)	&	-13(5)	&	14(5)	&	0.76(11)	\\
		48.11	&	73(8)	&	75(7)	&	13(1)	&	1.17(2)	&		&		&		&		&		\\
		\bottomrule
	\end{tabular}}
\end{table}

%%%%%%%%%%%%%%%%%%%%%%%%%%%%%%%%%%%%%%%%%%Spot migration

%\subsection{}

\section{Combination frequencies}
\label{sec:App1}
Table~\ref{tab:DepFreqKIC085} contains the values of the depended frequencies $f_{\rm i}$ (where $i$ is an increasing number), semi-amplitudes $A$, phases $\Phi$, and S/N for KIC~8504570. Moreover, in the last column of this table, the most possible combination for each frequency is also given. The combinations were calculated only for the first 255 frequencies because this is the maximum number of frequencies that the software can detect during one run (i.e. Fourier model). However, it should be noted that, in order to continue the search, the residuals from these Fourier models were given as new data sets to the same software, but no combinations could be calculated using the first 255 frequencies.

%%%%%%%%%%%%%%%%%%%%%%%%%%%%%%%%%%%%%%%%%%%%%%%%%%%%%%%%%%%%%%%%%%%%%%%%%%%%%%%%%%%%%%%%%% Table for Depedent Freqs for KIC085 %%%%%%%%%%%%%%%%%%%%%%%%%%%%%%%%%%%%%%%%%%%%%%%%%%%%%%%%%%%%
\begin{landscape}
	\begin{table}			
		\centering
		\caption{Combination frequencies of KIC~8504570.}
		\label{tab:DepFreqKIC085}
		%\scalebox{0.85}{
		\begin{tabular}{cc cc cl cc cc cl}
			\toprule																										
			$i$	&	$f_{\rm	i}$	&	$A$	&	$\Phi$	&	S/N	&	Combination				&	$i$	&	$f_{\rm	i}$	&	$A$	&	$\Phi$	&	S/N	&	Combination		\\
			&	(d$^{-1}$)	&	(mmag)	&	($^\circ$)	&	&						&		&	(d$^{-1}$)	&	(mmag)	&	($^\circ$)	&	&				\\
			\midrule
			5	&	13.96772(2)	&	1.462(5)	&	25.8(2)	&	194.5	&	$f_1-2f_{\rm orb}$	&	44	&	15.81073(12)	&	0.195(5)	&	172.1(1.3)	&	26	&	$f_{2}$+$f_{39}-f_{4}$	\\
			6	&	14.26173(2)	&	1.197(5)	&	127.1(2)	&	159.2	&	$f_1$+$2f_4-2f_5$	&	45	&	0.46827(12)	&	0.193(5)	&	315.2(1.4)	&	25.7	&	$2f_{12}$	\\
			7	&	13.87519(2)	&	1.187(5)	&	304.5(2)	&	158	&	$f_1$+$f_5-f_2$	&	46	&	14.30447(12)	&	0.191(5)	&	262.1(1.4)	&	25.4	&	$f_{3}-f_{8}$	\\
			9	&	14.15746(4)	&	0.620(5)	&	84.7(4)	&	82.6	&	$f_1$+$f_6-f_2$	&	47	&	18.43246(13)	&	0.190(5)	&	331.8(1.4)	&	25.3	&	$f_{18}$+$f_{5}-f_{1}$	\\
			11	&	14.52475(4)	&	0.556(5)	&	315.2(5)	&	74	&	$f_2$+$f_5-f_4$	&	48	&	18.21829(13)	&	0.188(5)	&	159.1(1.4)	&	25	&	$f_{37}-f_{39}$	\\
			12	&	0.23719(4)	&	0.539(5)	&	127.8(5)	&	71.7	&	$f_9-f_4$	&	49	&	10.27324(13)	&	0.184(5)	&	131.9(1.4)	&	24.5	&	$f_{21}-f_{2}$	\\
			13	&	25.59738(5)	&	0.467(5)	&	196.9(6)	&	62.1	&	$f_3$+$f_7-f_2$	&	50	&	25.41796(13)	&	0.180(5)	&	185.4(1.5)	&	23.9	&	$f_{2}$+$f_{22}-f_{6}$	\\
			14	&	27.16046(5)	&	0.469(5)	&	134.7(6)	&	62.3	&	$3f_7-f_2$	&	51	&	13.41866(14)	&	0.176(5)	&	265.4(1.5)	&	23.4	&	$f_{4}-2f_{\rm orb}$	\\
			15	&	30.46276(5)	&	0.464(5)	&	153.0(6)	&	61.7	&	$f_{11}$+$2f_4-f_8$	&	52	&	0.28509(14)	&	0.175(5)	&	333.8(1.5)	&	23.3	&	$f_{6}-f_{5}$	\\
			16	&	27.61182(5)	&	0.517(5)	&	4.5(5)	&	68.9	&	$f_1$+$f_{14}-f_4$	&	53	&	32.03430(14)	&	0.173(5)	&	177.3(1.5)	&	23	&	$f_{21}$+$f_{41}$	\\
			17	&	27.11303(5)	&	0.462(5)	&	323.8(6)	&	61.5	&	$f_{16}$+$f_{5}-f_2$	&	54	&	1.67956(14)	&	0.169(5)	&	153.5(1.6)	&	22.5	&	$f_{4}$+$f_{5}-f_3$	\\
			18	&	18.83497(5)	&	0.441(5)	&	315.3(6)	&	58.7	&	$f_{15}$+$f_{5}-f_{13}$	&	55	&	36.13223(14)	&	0.166(5)	&	97.7(1.6)	&	22.1	&	$\sim f_{31}$	\\
			19	&	0.49692(6)	&	0.388(5)	&	245.8(7)	&	51.6	&	$2f_{\rm orb}$	&	56	&	34.65322(14)	&	0.166(5)	&	239.4(1.6)	&	22.2	&	$f_{25}$+$f_{28}$	\\
			20	&	21.93578(6)	&	0.411(5)	&	94.0(6)	&	54.7	&	$2f_3-f_{15}$	&	57	&	32.10851(14)	&	0.167(5)	&	152.4(1.6)	&	22.2	&	$f_{10}$+$f_{46}$	\\
			21	&	24.73787(6)	&	0.409(5)	&	44.4(6)	&	54.4	&	$f_{17}$+$f_{8}-f_{6}$	&	58	&	22.50879(14)	&	0.166(5)	&	266.8(1.6)	&	22.1	&	$f_{34}-f_{52}$	\\
			22	&	25.20896(6)	&	0.402(5)	&	29.2(7)	&	53.5	&	$f_{13}$+$f_{7}-f_{6}$	&	59	&	36.09231(15)	&	0.163(5)	&	251.8(1.6)	&	21.7	&	$f_{20}$+$f_{9}$	\\
			23	&	27.65926(6)	&	0.386(5)	&	351.2(7)	&	51.4	&	$f_{14}$+$2f_{\rm orb}$	&	60	&	12.96918(15)	&	0.162(5)	&	42.6(1.6)	&	21.5	&	$f_{17}-f_{9}$	\\
			24	&	28.33090(6)	&	0.370(5)	&	248.9(7)	&	49.2	&	$f_{1}$+$f_{5}$	&	61	&	27.46012(15)	&	0.162(5)	&	65.2(1.6)	&	21.5	&	$f_{40}$+$f_{60}$	\\
			25	&	13.44778(7)	&	0.354(5)	&	305.5(7)	&	47.1	&	$f_{16}-f_{9}$	&	62	&	15.58199(15)	&	0.161(5)	&	230.0(1.6)	&	21.4	&	$f_{61}-f_{8}$	\\
			26	&	24.70969(7)	&	0.347(5)	&	0.6(8)	&	46.1	&	$f_{22}-2f_{\rm orb}$	&	63	&	29.89821(15)	&	0.160(5)	&	223.4(1.7)	&	21.2	&	$f_{46}$+$f_{62}$	\\
			27	&	13.99402(7)	&	0.341(5)	&	100.8(8)	&	45.3	&	$f_1$+$f_{7}-f_6$	&	64	&	13.69577(15)	&	0.157(5)	&	334.2(1.7)	&	20.9	&	$f_{16}-f_{4}$	\\
			28	&	21.19886(8)	&	0.306(5)	&	75.1(9)	&	40.7	&	$f_{18}$+$f_{6}-f_8$	&	65	&	29.36747(15)	&	0.155(5)	&	212.2(1.7)	&	20.7	&	$f_{42}$+$f_{8}$	\\
			29	&	24.98915(8)	&	0.284(5)	&	211.7(9)	&	37.8	&	$f_{22}$+$f_{9}-f_1$	&	66	&	0.26866(16)	&	0.152(5)	&	22.9(1.7)	&	20.3	&	$\sim f_{\rm orb}$	\\
			30	&	17.78666(9)	&	0.276(5)	&	127.0(1.0)	&	36.7	&	$f_{10}$+$f_{5}-f_{27}$	&	67	&	36.62210(16)	&	0.152(5)	&	336.5(1.7)	&	20.3	&	$f_{21}$+$f_{8}$	\\
			31	&	36.12330(10)	&	0.240(5)	&	139.7(1.1)	&	32	&	$f_{3}$+$f_{29}-f_{2}$	&	68	&	25.23620(16)	&	0.148(5)	&	271.9(1.8)	&	19.7	&	$f_{21}$+$2f_{\rm orb}$	\\
			32	&	36.06506(10)	&	0.232(5)	&	18.6(1.1)	&	30.8	&	$f_{1}$+$f_{20}-f_{12}$	&	69	&	1.75471(17)	&	0.143(5)	&	195.6(1.8)	&	19	&	$7f_{\rm orb}$	\\
			33	&	7.06065(10)	&	0.231(5)	&	274.2(1.1)	&	30.7	&	$f_{1}$+$f_{7}-f_{28}$	&	70	&	23.00711(17)	&	0.138(5)	&	131.6(1.9)	&	18.4	&	$f_{58}$+$2f_{\rm orb}$	\\
			34	&	22.78777(10)	&	0.229(5)	&	349.1(1.1)	&	30.5	&	$f_{1}$+$f_{3}-f_{30}$	&	71	&	14.60554(17)	&	0.137(5)	&	235.0(1.9)	&	18.2	&	$2f_{41}$	\\
			35	&	36.56292(10)	&	0.228(5)	&	75.9(1.2)	&	30.3	&	$2f_{\rm orb}$+$f_{32}$	&	72	&	0.24564(18)	&	0.135(5)	&	61.3(1.9)	&	18	&	$\sim f_{12}$	\\
			36	&	0.48424(11)	&	0.227(5)	&	125.2(1.2)	&	30.2	&	$\sim 2f_{\rm orb}$	&	73	&	14.41579(18)	&	0.135(5)	&	140.8(1.9)	&	18	&	$f_{4}$+$2f_{\rm orb}$	\\
			37	&	33.50439(11)	&	0.223(5)	&	66.3(1.2)	&	29.7	&	$f_{1}$+$f_{3}-f_{33}$	&	74	&	30.07058(18)	&	0.135(5)	&	341.8(2.0)	&	17.9	&	$f_{44}$+$f_{6}$	\\
			38	&	28.75971(11)	&	0.219(5)	&	74.6(1.2)	&	29.1	&	$2f_1$	&	75	&	36.59298(18)	&	0.132(5)	&	243.6(2.0)	&	17.5	&	$f_{26}$+$f_{8}$	\\
			39	&	15.27483(11)	&	0.212(5)	&	334.8(1.2)	&	28.2	&	$f_{14}-f_{8}$	&	76	&	36.14820(18)	&	0.133(5)	&	101.8(2.0)	&	17.7	&	$f_{67}-f_{36}$	\\
			40	&	14.49281(12)	&	0.206(5)	&	320.2(1.3)	&	27.4	&	$f_{12}$+$f_{6}$	&	77	&	1.61146(19)	&	0.129(5)	&	94.4(2.0)	&	17.1	&	$f_{62}-f_{5}$	\\
			41	&	7.30582(12)	&	0.202(5)	&	98.9(1.3)	&	26.9	&	$f_{1}-f_{33}$	&	78	&	11.09517(18)	&	0.130(5)	&	327.0(2.0)	&	17.3	&	$f_{13}-f_{40}$	\\
			42	&	17.46822(12)	&	0.199(5)	&	106.4(1.3)	&	26.5	&	$f_{10}$+$f_{4}-f_{6}$	&	79	&	29.86627(19)	&	0.129(5)	&	249.8(2.0)	&	17.1	&	$f_{65}$+$2f_{\rm orb}$	\\
			43	&	0.25832(12)	&	0.199(5)	&	80.8(1.3)	&	26.4	&	$f_{\rm orb}$	&	80	&	29.39894(19)	&	0.128(5)	&	19.2(2.1)	&	17	&	$f_{63}-2f_{\rm orb}$	\\
			\bottomrule																									
		\end{tabular}%}																											
	\end{table}																									\end{landscape}

\begin{landscape}
	\begin{table}		
		\centering																											
		%\contcaption{   }																												
		%\scalebox{0.90}{																												
		\begin{tabular}{cc cc cl cc cc cl}																												
		\toprule																											
			$i$	&	$f_{\rm	i}$	&	$A$	&	$\Phi$	&	S/N	&	Combination				&	$i$	&	$f_{\rm	i}$	&	$A$	&	$\Phi$	&	S/N	&	Combination		\\
			&	(d$^{-1}$)	&	(mmag)	&	($^\circ$)	&	&						&		&	(d$^{-1}$)	&	(mmag)	&	($^\circ$)	&	&				\\
			\midrule
			81	&	29.65398(19)	&	0.128(5)	&	5.7(2.1)	&	17	&	$f_{1}$+$f_{39}$	&	118	&	1.88387(27)	&	0.088(5)	&	75.6(3.0)	&	11.7	&	$f_{44}-f_{4}$	\\
			82	&	0.50819(19)	&	0.127(5)	&	319.2(2.1)	&	16.9	&	$\sim 2f_{\rm orb}$	&	119	&	23.64493(27)	&	0.088(5)	&	357.2(3.0)	&	11.7	&	$f_{67}-f_{60}$	\\
			83	&	13.62955(19)	&	0.125(5)	&	139.7(2.1)	&	16.6	&	$f_{7}-f_{12}$	&	120	&	23.41808(27)	&	0.087(5)	&	79.5(3.0)	&	11.6	&	$f_{119}-f_{12}$	\\
			84	&	31.77175(20)	&	0.122(5)	&	241.8(2.2)	&	16.2	&	$f_{10}$+$f_{5}$	&	121	&	0.59273(27)	&	0.087(5)	&	226.5(3.0)	&	11.5	&	$f_{2}-f_{7}$	\\
			85	&	26.41931(20)	&	0.121(5)	&	35.6(2.2)	&	16.1	&	$f_{12}$+$f_{3}$	&	122	&	0.45887(27)	&	0.088(5)	&	318.0(3.0)	&	11.7	&	$\sim f_{45}$	\\
			86	&	16.23344(20)	&	0.119(5)	&	170.1(2.2)	&	15.8	&	$f_{2}$+$f_{69}$	&	123	&	0.21840(27)	&	0.088(5)	&	216.5(3.0)	&	11.7	&	$f_{2}-f_{6}$	\\
			87	&	0.82146(20)	&	0.117(5)	&	183.8(2.2)	&	15.6	&	$f_{39}-f_{2}$	&	124	&	27.10128(28)	&	0.086(5)	&	248.2(3.1)	&	11.5	&	$\sim f_{17}$	\\
			88	&	13.94799(20)	&	0.117(5)	&	250.4(2.3)	&	15.6	&	$f_{24}-f_{1}$	&	125	&	0.03147(28)	&	0.086(5)	&	239.7(3.1)	&	11.4	&	$f_{4}-f_{7}$	\\
			89	&	27.26003(20)	&	0.117(5)	&	220.8(2.3)	&	15.6	&	$2f_{83}$	&	126	&	0.14043(28)	&	0.087(5)	&	63.0(3.0)	&	11.5	&	$f_{11}-f_{1}$	\\
			90	&	16.75243(21)	&	0.116(5)	&	61.5(2.3)	&	15.4	&	$f_{53}-f_{39}$	&	127	&	34.64194(28)	&	0.086(5)	&	264.6(3.1)	&	11.4	&	$\sim f_{56}$	\\
			91	&	15.05783(21)	&	0.113(5)	&	108.8(2.3)	&	15.1	&	$f_{25}$+$f_{77}$	&	128	&	0.94076(28)	&	0.084(5)	&	79.9(3.1)	&	11.2	&	$2f_{45}$	\\
			92	&	0.52933(21)	&	0.113(5)	&	141.6(2.3)	&	15.1	&	$2f_{\rm orb}$	&	129	&	36.05708(28)	&	0.084(5)	&	156.5(3.1)	&	11.2	&	$\sim f_{32}$	\\
			93	&	0.34709(21)	&	0.112(5)	&	44.7(2.3)	&	15	&	$f_{6}-f_{4}$	&	130	&	10.15488(29)	&	0.083(5)	&	350.0(3.2)	&	11.1	&	$f_{16}-f_{42}$	\\
			94	&	18.74761(21)	&	0.111(5)	&	155.5(2.4)	&	14.8	&	$f_{35}-f_{10}$	&	131	&	18.94535(29)	&	0.083(5)	&	120.1(3.2)	&	11.1	&	$f_{33}$+$f_{8}$	\\
			95	&	36.63478(22)	&	0.108(5)	&	271.3(2.4)	&	14.4	&	$\sim f_{67}$	&	132	&	31.73511(29)	&	0.083(5)	&	308.6(3.2)	&	11	&	$f_{10}$+$f_{4}$	\\
			96	&	12.27641(22)	&	0.109(5)	&	16.7(2.4)	&	14.4	&	$f_{3}-f_{4}$	&	133	&	0.73974(29)	&	0.082(5)	&	202.6(3.2)	&	11	&	$f_{1}-f_{83}$	\\
			97	&	22.96202(22)	&	0.106(5)	&	296.7(2.5)	&	14.1	&	$f_{26}-f_{69}$	&	134	&	16.30624(30)	&	0.080(5)	&	50.4(3.3)	&	10.7	&	$f_{15}-f_{9}$	\\
			98	&	22.46229(22)	&	0.109(5)	&	292.1(2.4)	&	14.5	&	$f_{67}-f_{9}$	&	135	&	16.02725(30)	&	0.080(5)	&	134.9(3.3)	&	10.7	&	$f_{63}-f_{7}$	\\
			99	&	21.69765(23)	&	0.105(5)	&	288.0(2.5)	&	13.9	&	$f_{32}-f_{1}$	&	136	&	31.16305(30)	&	0.080(5)	&	85.4(3.3)	&	10.6	&	$2f_{62}$	\\
			100	&	4.37269(23)	&	0.103(5)	&	131.4(2.5)	&	13.8	&	$f_{18}-f_{2}$	&	137	&	12.97670(30)	&	0.080(5)	&	177.2(3.3)	&	10.6	&	$\sim f_{60}$	\\
			101	&	21.88975(24)	&	0.101(5)	&	143.3(2.6)	&	13.4	&	$f_{76}-f_{6}$	&	138	&	0.99431(30)	&	0.080(5)	&	127.4(3.3)	&	10.6	&	$4f_{\rm orb}$	\\
			102	&	13.65820(24)	&	0.100(5)	&	55.8(2.6)	&	13.3	&	$f_{4}-f_{\rm orb}$	&	139	&	27.62122(30)	&	0.079(5)	&	208.7(3.3)	&	10.6	&	$\sim f_{16}$	\\
			103	&	36.64746(24)	&	0.099(5)	&	305.0(2.7)	&	13.2	&	$\sim f_{95}$	&	140	&	10.03276(30)	&	0.079(5)	&	321.0(3.3)	&	10.5	&	$f_{20}-f_{8}$	\\
			104	&	27.34082(24)	&	0.099(5)	&	147.9(2.7)	&	13.1	&	$f_{1}$+$f_{60}$	&	141	&	38.46793(31)	&	0.078(5)	&	45.6(3.4)	&	10.3	&	$f_{3}$+$f_{96}$	\\
			105	&	26.63865(24)	&	0.099(5)	&	96.3(2.6)	&	13.2	&	$f_{1}$+$f_{96}$	&	142	&	36.07352(31)	&	0.078(5)	&	157.9(3.4)	&	10.3	&	$\sim f_{32}$	\\
			106	&	32.23532(25)	&	0.097(5)	&	127.1(2.7)	&	12.9	&	$f_{10}$+$f_{73}$	&	143	&	21.94752(31)	&	0.078(5)	&	0.3(3.4)	&	10.3	&	$\sim f_{20}$	\\
			107	&	36.11673(25)	&	0.097(5)	&	107.4(2.7)	&	12.9	&	$\sim f_{31}$	&	144	&	26.66871(31)	&	0.077(5)	&	44.1(3.4)	&	10.3	&	$f_{3}$+$f_{36}$	\\
			108	&	36.57325(25)	&	0.096(5)	&	208.1(2.7)	&	12.8	&	$\sim f_{35}$	&	145	&	18.99466(31)	&	0.076(5)	&	169.6(3.5)	&	10.1	&	$f_{12}$+$f_{94}$	\\
			109	&	32.22593(25)	&	0.095(5)	&	65.5(2.8)	&	12.6	&	$\sim f_{106}$	&	146	&	13.85546(31)	&	0.076(5)	&	39.8(3.5)	&	10.1	&	$f_{24}-f_{2}$	\\
			110	&	12.35531(25)	&	0.095(5)	&	281.4(2.8)	&	12.6	&	$f_{45}$+$f_{8}$	&	147	&	14.35473(31)	&	0.078(5)	&	76.3(3.4)	&	10.4	&	$f_{24}-f_{5}$	\\
			111	&	0.47672(25)	&	0.094(5)	&	33.8(2.8)	&	12.5	&	$\sim f_{36}$	&	148	&	21.47033(32)	&	0.075(5)	&	150.6(3.5)	&	10	&	$f_{41}$+$f_{9}$	\\
			112	&	11.69260(26)	&	0.092(5)	&	19.3(2.9)	&	12.2	&	$f_{13}-f_{4}$	&	149	&	26.14173(32)	&	0.076(5)	&	14.0(3.5)	&	10.1	&	$f_{6}$+$f_{8}$	\\
			113	&	21.44966(26)	&	0.092(5)	&	127.2(2.9)	&	12.2	&	$f_{12}$+$f_{28}$	&	150	&	25.03236(32)	&	0.075(5)	&	283.6(3.5)	&	9.9	&	$f_{21}$+$f_{52}$	\\
			114	&	36.55494(26)	&	0.091(5)	&	211.4(2.9)	&	12.1	&	$\sim f_{35}$	&	151	&	28.36237(32)	&	0.074(5)	&	211.9(3.5)	&	9.9	&	$f_{1}$+$f_{27}$	\\
			115	&	22.08279(26)	&	0.090(5)	&	140.7(2.9)	&	12	&	$f_{35}-f_{2}$	&	152	&	21.67229(32)	&	0.074(5)	&	20.0(3.6)	&	9.9	&	$f_{41}$+$f_{1}$	\\
			116	&	29.08191(27)	&	0.089(5)	&	203.6(2.9)	&	11.9	&	$f_{71}$+$f_{2}$	&	153	&	10.95004(32)	&	0.074(5)	&	127.8(3.6)	&	9.8	&	$f_{22}-f_{6}$	\\
			117	&	30.36789(27)	&	0.089(5)	&	120.1(2.9)	&	11.9	&	$f_{79}$+$2f_{\rm orb}$	&	154	&	31.12219(33)	&	0.072(5)	&	30.7(3.6)	&	9.6	&	$f_{1}$+$f_{90}$	\\
			\bottomrule																									
		\end{tabular}%}																												
	\end{table}																												
\end{landscape}

\begin{landscape}
	\begin{table}		
		\centering																											
		%\contcaption{   }																												
		%\scalebox{0.90}{																												
		\begin{tabular}{cc cc cl cc cc cl}																												
			\toprule																											
			$i$	&	$f_{\rm	i}$	&	$A$	&	$\Phi$	&	S/N	&	Combination				&	$i$	&	$f_{\rm	i}$	&	$A$	&	$\Phi$	&	S/N	&	Combination		\\
			&	(d$^{-1}$)	&	(mmag)	&	($^\circ$)	&	&						&		&	(d$^{-1}$)	&	(mmag)	&	($^\circ$)	&	&				\\
			\midrule
			155	&	0.72753(33)	&	0.072(5)	&	11.2(3.6)	&	9.6	&	$\sim f_{133}$	&	192	&	40.56879(40)	&	0.060(5)	&	122.9(4.4)	&	7.9	&	$f_{3}$+$f_{1}$	\\
			156	&	17.96654(33)	&	0.072(5)	&	292.0(3.7)	&	9.6	&	$f_{79}-f_{8}$	&	193	&	23.34011(40)	&	0.059(5)	&	104.6(4.4)	&	7.9	&	$f_{163}-f_{11}$	\\
			157	&	31.72666(33)	&	0.072(5)	&	225.9(3.7)	&	9.5	&	$\sim f_{132}$	&	194	&	24.53309(40)	&	0.059(5)	&	54.8(4.5)	&	7.8	&	$f_{49}$+$f_{6}$	\\
			158	&	31.36642(33)	&	0.072(5)	&	21.3(3.7)	&	9.6	&	$f_{2}$+$f_{154}$	&	195	&	0.74773(41)	&	0.059(5)	&	131.1(4.5)	&	7.8	&	$\sim f_{133}$	\\
			159	&	26.18870(34)	&	0.071(5)	&	52.9(3.7)	&	9.4	&	$\sim f_{3}$	&	196	&	1.98861(41)	&	0.058(5)	&	212.3(4.5)	&	7.7	&	$f_{7}-f_{8}$	\\
			160	&	41.07370(34)	&	0.070(5)	&	250.4(3.8)	&	9.3	&	$f_{14}$+$f_{4}$	&	197	&	14.38056(41)	&	0.058(5)	&	183.0(4.5)	&	7.7	&	$\sim f_{1}$	\\
			161	&	16.77450(34)	&	0.069(5)	&	135.3(3.8)	&	9.2	&	$f_{15}-f_{64}$	&	198	&	41.29397(41)	&	0.058(5)	&	78.0(4.6)	&	7.7	&	$f_{80}$+$f_{8}$	\\
			162	&	33.51284(35)	&	0.069(5)	&	163.5(3.8)	&	9.2	&	$\sim f_{37}$	&	199	&	0.06059(42)	&	0.057(5)	&	1.5(4.6)	&	7.6	&	$f_{5}-f_{4}$	\\
			163	&	37.86299(35)	&	0.069(5)	&	171.6(3.8)	&	9.1	&	$f_{13}$+$f_{96}$	&	200	&	13.73006(42)	&	0.057(5)	&	265.5(4.7)	&	7.5	&	$f_{5}-f_{12}$	\\
			164	&	36.10311(35)	&	0.068(5)	&	194.3(3.9)	&	9.1	&	$\sim f_{59}$	&	201	&	25.09952(42)	&	0.056(5)	&	142.7(4.7)	&	7.5	&	$f_{13}-2f_{\rm orb}$	\\
			165	&	37.85124(35)	&	0.068(5)	&	324.1(3.9)	&	9	&	$\sim f_{163}$	&	202	&	27.83304(42)	&	0.057(5)	&	192.4(4.7)	&	7.5	&	$2f_{4}$	\\
			166	&	36.03547(36)	&	0.067(5)	&	275.4(3.9)	&	8.9	&	$f_{1}$+$f_{152}$	&	203	&	0.21229(43)	&	0.056(5)	&	71.7(4.7)	&	7.4	&	$\sim f_{123}$	\\
			167	&	36.53427(34)	&	0.070(5)	&	313.3(3.8)	&	9.3	&	$f_{71}$+$f_{20}$	&	204	&	13.37545(44)	&	0.055(5)	&	311.6(4.8)	&	7.3	&	$f_{7}-2f_{\rm orb}$	\\
			168	&	27.15624(36)	&	0.067(5)	&	43.3(3.9)	&	8.9	&	$\sim f_{14}$	&	205	&	27.87061(44)	&	0.055(5)	&	247.7(4.8)	&	7.3	&	$f_{7}$+$f_{27}$	\\
			169	&	27.71186(36)	&	0.066(5)	&	241.6(4.0)	&	8.8	&	$2f_{146}$	&	206	&	36.62821(44)	&	0.055(5)	&	312.0(4.8)	&	7.3	&	$\sim f_{67}$	\\
			170	&	27.60008(36)	&	0.066(5)	&	276.2(4.0)	&	8.8	&	$\sim f_{16}$	&	207	&	40.66602(44)	&	0.054(5)	&	243.5(4.9)	&	7.2	&	$f_{3}$+$f_{2}$	\\
			171	&	18.39489(36)	&	0.066(5)	&	195.7(4.0)	&	8.8	&	$f_{10}$+$f_{121}$	&	208	&	32.08314(44)	&	0.054(5)	&	324.2(4.9)	&	7.2	&	$f_{7}$+$f_{48}$	\\
			172	&	0.12212(36)	&	0.066(5)	&	320.6(4.0)	&	8.8	&	$f_{1}-f_{6}$	&	209	&	34.06565(43)	&	0.055(5)	&	166.9(4.8)	&	7.3	&	$f_{56}$+$f_{121}$	\\
			173	&	26.16757(37)	&	0.065(5)	&	250.0(4.0)	&	8.7	&	$f_{2}$+$f_{112}$	&	210	&	0.53590(44)	&	0.054(5)	&	260.0(4.9)	&	7.1	&	$\sim f_{92}$	\\
			174	&	0.16955(37)	&	0.065(5)	&	161.0(4.1)	&	8.6	&	$f_{2}-f_{46}$	&	211	&	13.27118(45)	&	0.054(5)	&	98.1(4.9)	&	7.1	&	$f_{23}-f_{1}$	\\
			175	&	1.45412(37)	&	0.064(5)	&	238.5(4.1)	&	8.6	&	$f_{3}-f_{21}$	&	212	&	36.65545(45)	&	0.053(5)	&	295.0(4.9)	&	7.1	&	$\sim f_{103}$	\\
			176	&	34.00412(37)	&	0.064(5)	&	68.4(4.1)	&	8.5	&	$2f_{\rm orb}$+$f_{37}$	&	213	&	19.21306(45)	&	0.053(5)	&	121.4(4.9)	&	7.1	&	$f_{37}-f_{46}$	\\
			177	&	14.17625(38)	&	0.063(5)	&	215.8(4.2)	&	8.4	&	$f_{2}-f_{52}$	&	214	&	17.64669(45)	&	0.053(5)	&	359.8(5.0)	&	7.0	&	$f_{53}-f_{1}$	\\
			178	&	31.61018(38)	&	0.063(5)	&	119.0(4.2)	&	8.4	&	$2f_{44}$	&	215	&	24.26679(45)	&	0.053(5)	&	326.2(5.2)	&	7.0	&	$f_{76}-f_{8}$	\\
			179	&	31.57402(37)	&	0.064(5)	&	155.1(4.1)	&	8.6	&	$f_{53}-f_{45}$	&	216	&	0.90554(46)	&	0.052(5)	&	216.3(5.1)	&	6.9	&	$f_{17}-f_{3}$	\\
			180	&	0.20196(38)	&	0.063(5)	&	242.0(4.2)	&	8.4	&	$f_{2}-f_{6}$	&	217	&	18.97682(46)	&	0.052(5)	&	3.4(5.1)	&	6.9	&	$f_{37}-f_{11}$	\\
			181	&	0.29824(34)	&	0.070(5)	&	324.0(3.8)	&	9.3	&	$\sim f_{52}$	&	218	&	25.70775(46)	&	0.052(5)	&	220.8(5.1)	&	6.9	&	$f_{3}-2f_{\rm orb}$	\\
			182	&	18.49634(38)	&	0.062(5)	&	290.9(4.2)	&	8.3	&	$f_{18}-f_{93}$	&	219	&	41.04270(46)	&	0.052(5)	&	92.9(5.1)	&	6.9	&	$f_{17}$+$f_{4}$	\\
			183	&	25.20520(38)	&	0.062(5)	&	94.1(4.2)	&	8.3	&	$\sim f_{22}$	&	220	&	0.08877(46)	&	0.052(5)	&	110.0(5.1)	&	6.9	&	$f_{2}-f_{1}$	\\
			184	&	28.10451(39)	&	0.062(5)	&	305.5(4.3)	&	8.2	&	$f_{2}$+$f_{83}$	&	221	&	3.05619(46)	&	0.052(5)	&	32.8(5.1)	&	6.9	&	$f_{29}-f_{20}$	\\
			185	&	30.26831(39)	&	0.062(5)	&	199.9(4.3)	&	8.2	&	$f_{2}$+$f_{44}$	&	222	&	14.02595(46)	&	0.051(5)	&	47.5(5.1)	&	6.8	&	$f_{6}-f_{12}$	\\
			186	&	16.08924(39)	&	0.062(5)	&	146.7(4.3)	&	8.2	&	$f_{15}-f_{1}$	&	223	&	10.47896(47)	&	0.051(5)	&	73.4(5.2)	&	6.8	&	$f_{21}-f_{6}$	\\
			187	&	3.65737(39)	&	0.061(5)	&	46.8(4.3)	&	8.2	&	$f_{10}-f_{9}$	&	224	&	30.62339(47)	&	0.051(5)	&	314.4(5.2)	&	6.8	&	$f_{7}$+$f_{90}$	\\
			188	&	3.40985(38)	&	0.062(5)	&	305.7(4.3)	&	8.2	&	$f_{30}-f_{1}$	&	225	&	22.68914(47)	&	0.051(5)	&	177.7(5.2)	&	6.8	&	$f_{35}-f_{7}$	\\
			189	&	17.81296(39)	&	0.061(5)	&	136.4(4.3)	&	8.2	&	$\sim f_{10}$	&	226	&	7.57260(47)	&	0.051(5)	&	39.1(5.2)	&	6.7	&	$f_{20}-f_{1}$	\\
			190	&	28.39102(40)	&	0.060(5)	&	214.7(4.4)	&	8	&	$f_{2}$+$f_{4}$	&	227	&	38.64171(47)	&	0.050(5)	&	98.2(5.2)	&	6.7	&	$f_{26}$+$f_{41}$	\\
			191	&	34.45924(40)	&	0.060(5)	&	2.7(4.4)	&	7.9	&	$f_{14}$+$f_{41}$	&	228	&	38.14057(42)	&	0.057(5)	&	102.1(4.7)	&	7.5	&	$f_{215}$+$f_{7}$	\\
			\bottomrule																								
		\end{tabular}%}																												
	\end{table}																							\end{landscape}

\begin{landscape}			
	\begin{table}		
		\centering																											
		%\contcaption{   }																												
		%\scalebox{0.90}{																												
		\begin{tabular}{cc cc cl cc cc cl}																												
			\toprule																											
			$i$	&	$f_{\rm	i}$	&	$A$	&	$\Phi$	&	S/N	&	Combination				&	$i$	&	$f_{\rm	i}$	&	$A$	&	$\Phi$	&	S/N	&	Combination		\\
			&	(d$^{-1}$)	&	(mmag)	&	($^\circ$)	&	&						&		&	(d$^{-1}$)	&	(mmag)	&	($^\circ$)	&	&				\\
			\midrule
			229	&	1.59878(48)	&	0.050(5)	&	8.5(5.3)	&	6.7	&	$\sim f_{77}$	&	266	&	23.07991(50)	&	0.043(4)	&	337.3(5.5)	&	5.7	&		\\
			230	&	27.08297(48)	&	0.050(5)	&	107.0(5.3)	&	6.6	&	$f_{219}-f_{5}$	&	267	&	32.05825(50)	&	0.042(4)	&	235.4(5.6)	&	5.6	&		\\
			231	&	0.70827(48)	&	0.049(5)	&	220.6(5.3)	&	6.6	&	$f_{1}-f_{102}$	&	268	&	38.92399(50)	&	0.042(4)	&	157.0(5.6)	&	5.6	&		\\
			232	&	34.62597(49)	&	0.048(5)	&	111.9(5.4)	&	6.5	&	$f_{51}$+$f_{28}$	&	269	&	13.99214(50)	&	0.042(4)	&	310.0(5.6)	&	5.6	&		\\
			233	&	0.38842(49)	&	0.048(5)	&	314.4(5.4)	&	6.4	&	$f_{6}-f_{7}$	&	270	&	31.07569(51)	&	0.042(4)	&	107.8(5.6)	&	5.6	&		\\
			234	&	15.59984(49)	&	0.048(5)	&	73.3(5.5)	&	6.4	&	$f_{74}-f_{2}$	&	271	&	31.39131(50)	&	0.042(4)	&	171.7(5.6)	&	5.6	&		\\
			235	&	40.92481(50)	&	0.048(5)	&	116.2(5.5)	&	6.4	&	$f_{6}$+$f_{144}$	&	272	&	0.12822(51)	&	0.042(4)	&	86.1(5.6)	&	5.6	&		\\
			236	&	24.74257(50)	&	0.048(5)	&	157.2(5.5)	&	6.4	&	$\sim f_{21}$	&	273	&	18.60718(51)	&	0.042(4)	&	318.8(5.6)	&	5.6	&		\\
			237	&	0.19398(50)	&	0.047(5)	&	117.9(5.6)	&	6.3	&	$\sim f_{180}$	&	274	&	18.47614(50)	&	0.042(4)	&	353.0(5.6)	&	5.6	&		\\
			238	&	26.20326(50)	&	0.047(5)	&	297.0(5.6)	&	6.3	&	$\sim f_{3}$	&	275	&	40.52840(51)	&	0.042(4)	&	229.9(5.6)	&	5.6	&		\\
			239	&	0.83696(51)	&	0.047(5)	&	323.3(5.6)	&	6.3	&	$f_{2}-f_{83}$	&	276	&	18.33665(51)	&	0.042(4)	&	287.7(5.7)	&	5.5	&		\\
			240	&	0.86374(49)	&	0.049(5)	&	336.6(5.4)	&	6.5	&	$f_{13}-f_{21}$	&	277	&	11.47702(51)	&	0.042(4)	&	147.2(5.7)	&	5.5	&		\\
			241	&	28.33935(51)	&	0.047(5)	&	186.2(5.6)	&	6.3	&	$\sim f_{24}$	&	278	&	26.46018(51)	&	0.041(4)	&	1.4(5.7)	&	5.5	&		\\
			242	&	30.09078(51)	&	0.047(5)	&	216.2(5.6)	&	6.2	&	$f_{12}-f_{79}$	&	279	&	21.75167(52)	&	0.041(4)	&	3.4(5.7)	&	5.5	&		\\
			243	&	36.37364(51)	&	0.047(5)	&	196.1(5.6)	&	6.2	&	$f_{58}$+$f_{7}$	&	280	&	14.12834(52)	&	0.041(4)	&	1.5(5.7)	&	5.5	&		\\
			244	&	27.16892(51)	&	0.047(5)	&	218.8(5.6)	&	6.2	&	$\sim f_{14}$	&	281	&	27.27976(52)	&	0.041(4)	&	6.7(5.7)	&	5.5	&		\\
			245	&	22.59802(51)	&	0.047(5)	&	183.6(5.6)	&	6.2	&	$f_{35}-f_{5}$	&	282	&	52.44645(52)	&	0.041(4)	&	116.8(5.8)	&	5.4	&		\\
			246	&	41.05115(51)	&	0.046(5)	&	38.3(5.7)	&	6.2	&	$\sim f_{129}$	&	283	&	18.66448(53)	&	0.040(4)	&	330.3(5.8)	&	5.4	&		\\
			247	&	0.01738(52)	&	0.046(5)	&	358.1(5.7)	&	6.1	&	$f_{27}-f_{5}$	&	284	&	32.63830(53)	&	0.040(4)	&	272.2(5.8)	&	5.4	&		\\
			248	&	18.44796(52)	&	0.046(5)	&	233.8(5.7)	&	6.1	&	$f_{48}$+$f_{12}$	&	285	&	32.26773(53)	&	0.040(4)	&	144.4(5.8)	&	5.4	&		\\
			249	&	28.88230(53)	&	0.045(5)	&	324.7(5.8)	&	6	&	$f_{2}$+$f_{73}$	&	286	&	32.13246(53)	&	0.040(4)	&	58.5(5.8)	&	5.4	&		\\
			250	&	11.91476(53)	&	0.045(5)	&	207.0(5.8)	&	6	&	$f_{125}$+$f_{8}$	&	287	&	27.88330(53)	&	0.040(4)	&	75.1(5.9)	&	5.3	&		\\
			251	&	1.22304(53)	&	0.045(5)	&	218.1(5.9)	&	5.9	&	$f_{24}-f_{17}$	&	288	&	38.65768(53)	&	0.040(4)	&	300.7(5.9)	&	5.3	&		\\
			252	&	25.08167(54)	&	0.044(5)	&	194.0(5.9)	&	5.9	&	$f_{13}-f_{82}$	&	289	&	17.62462(53)	&	0.040(4)	&	300.3(5.9)	&	5.3	&		\\
			253	&	18.41978(54)	&	0.044(5)	&	313.2(6.0)	&	5.8	&	$\sim f_{47}$	&	290	&	34.28546(54)	&	0.040(4)	&	70.5(5.9)	&	5.3	&		\\
			254	&	0.41942(55)	&	0.044(5)	&	215.3(6.0)	&	5.8	&	$f_{1}-f_{5}$	&	291	&	0.57676(54)	&	0.040(4)	&	267.5(5.9)	&	5.3	&		\\
			255	&	38.19834(55)	&	0.044(5)	&	200.8(6.1)	&	5.8	&	$f_{21}$+$f_{25}$	&	292	&	3.91476(53)	&	0.040(4)	&	24.5(5.9)	&	5.3	&		\\
			256	&	37.87661(49)	&	0.044(4)	&	247.1(5.4)	&	5.8	&		&	293	&	0.18176(53)	&	0.040(4)	&	152.7(5.9)	&	5.3	&		\\
			257	&	31.67828(49)	&	0.043(4)	&	206.9(5.4)	&	5.8	&		&	294	&	40.69514(54)	&	0.040(4)	&	323.4(5.9)	&	5.3	&		\\
			258	&	40.59838(49)	&	0.043(4)	&	304.0(5.5)	&	5.7	&		&	295	&	29.56990(54)	&	0.039(4)	&	136.9(6.0)	&	5.2	&		\\
			259	&	21.17303(50)	&	0.043(4)	&	3.0(5.5)	&	5.7	&		&	296	&	30.29039(54)	&	0.039(4)	&	163.6(6.0)	&	5.3	&		\\
			260	&	0.78389(50)	&	0.043(4)	&	350.5(5.5)	&	5.7	&		&	297	&	27.21213(54)	&	0.039(4)	&	178.6(6.0)	&	5.2	&		\\
			261	&	2.67105(50)	&	0.043(4)	&	252.9(5.5)	&	5.7	&		&	298	&	0.98303(54)	&	0.039(4)	&	338.9(6.0)	&	5.2	&		\\
			262	&	5.59244(50)	&	0.043(4)	&	217.8(5.5)	&	5.7	&		&	299	&	36.43047(55)	&	0.039(4)	&	346.3(6.0)	&	5.2	&		\\
			263	&	14.36553(50)	&	0.043(4)	&	343.9(5.5)	&	5.7	&		&	300	&	36.04909(54)	&	0.039(4)	&	220.3(6.0)	&	5.2	&		\\
			264	&	26.13140(50)	&	0.042(4)	&	308.0(5.5)	&	5.6	&		&	301	&	0.49504(55)	&	0.039(4)	&	284.1(6.1)	&	5.1	&		\\
			265	&	22.87795(50)	&	0.042(4)	&	136.7(5.6)	&	5.6	&		&	302	&	22.02925(55)	&	0.039(4)	&	249.9(6.1)	&	5.1	&		\\
			\bottomrule																								
		\end{tabular}%}																												
	\end{table}																							\end{landscape}

\begin{landscape}			
	\begin{table}		
		\centering																											
		%\contcaption{   }																												
		%\scalebox{0.90}{																												
		\begin{tabular}{cc cc cl cc cc cl}																												
			\toprule																											
			$i$	&	$f_{\rm	i}$	&	$A$	&	$\Phi$	&	S/N	&	Combination				&	$i$	&	$f_{\rm	i}$	&	$A$	&	$\Phi$	&	S/N	&	Combination		\\
			&	(d$^{-1}$)	&	(mmag)	&	($^\circ$)	&	&						&		&	(d$^{-1}$)	&	(mmag)	&	($^\circ$)	&	&				\\
			\midrule
			303	&	22.84977(56)	&	0.038(4)	&	22.1(6.1)	&	5.1	&		&	340	&	2.33640(61)	&	0.035(4)	&	201.6(6.8)	&	4.6	&		\\
			304	&	38.82301(56)	&	0.038(4)	&	84.4(6.2)	&	5.1	&		&	341	&	5.79546(61)	&	0.035(4)	&	323.3(6.8)	&	4.6	&		\\
			305	&	36.08432(56)	&	0.038(4)	&	328.8(6.2)	&	5	&		&	342	&	8.99316(61)	&	0.035(4)	&	94.4(6.8)	&	4.6	&		\\
			306	&	17.28880(56)	&	0.038(4)	&	138.2(6.2)	&	5	&		&	343	&	40.57713(62)	&	0.035(4)	&	49.3(6.8)	&	4.6	&		\\
			307	&	32.83745(56)	&	0.038(4)	&	283.2(6.2)	&	5	&		&	344	&	23.28876(62)	&	0.034(4)	&	153.6(6.8)	&	4.6	&		\\
			308	&	0.44478(57)	&	0.038(4)	&	112.8(6.3)	&	5	&		&	345	&	32.32647(62)	&	0.034(4)	&	311.9(6.8)	&	4.6	&		\\
			309	&	0.40909(56)	&	0.038(4)	&	131.4(6.2)	&	5	&		&	346	&	38.62485(62)	&	0.034(4)	&	117.6(6.9)	&	4.5	&		\\
			310	&	20.29755(57)	&	0.038(4)	&	256.7(6.3)	&	5	&		&	347	&	43.33923(62)	&	0.034(4)	&	325.0(6.9)	&	4.5	&		\\
			311	&	25.51472(57)	&	0.037(4)	&	349.6(6.3)	&	5	&		&	348	&	26.17857(62)	&	0.034(4)	&	65.6(6.9)	&	4.5	&		\\
			312	&	34.44515(57)	&	0.037(4)	&	134.3(6.3)	&	5	&		&	349	&	24.85296(63)	&	0.034(4)	&	62.6(6.9)	&	4.5	&		\\
			313	&	31.94882(55)	&	0.038(4)	&	28.5(6.1)	&	5.1	&		&	350	&	1.30086(63)	&	0.034(4)	&	146.9(6.9)	&	4.5	&		\\
			314	&	34.08021(57)	&	0.038(4)	&	81.9(6.3)	&	5	&		&	351	&	12.47697(63)	&	0.034(4)	&	175.2(7.0)	&	4.5	&		\\
			315	&	22.79623(57)	&	0.037(4)	&	283.8(6.3)	&	5	&		&	352	&	22.24233(63)	&	0.034(4)	&	97.3(7.0)	&	4.5	&		\\
			316	&	13.47549(57)	&	0.037(4)	&	269.2(6.3)	&	4.9	&		&	353	&	13.96989(63)	&	0.034(4)	&	350.7(7.0)	&	4.5	&		\\
			317	&	15.35138(57)	&	0.037(4)	&	198.7(6.3)	&	4.9	&		&	354	&	0.11682(63)	&	0.034(4)	&	172.3(7.0)	&	4.5	&		\\
			318	&	26.50339(58)	&	0.037(4)	&	159.5(6.4)	&	4.9	&		&	355	&	38.84400(63)	&	0.034(4)	&	299.6(7.0)	&	4.5	&		\\
			319	&	14.42471(58)	&	0.037(4)	&	356.5(6.4)	&	4.9	&		&	356	&	36.01940(63)	&	0.034(4)	&	304.2(7.0)	&	4.5	&		\\
			320	&	18.24741(58)	&	0.037(4)	&	328.3(6.4)	&	4.9	&		&	357	&	0.37396(63)	&	0.034(4)	&	180.4(7.0)	&	4.5	&		\\
			321	&	17.79464(58)	&	0.037(4)	&	332.9(6.4)	&	4.9	&		&	358	&	41.03200(63)	&	0.034(4)	&	135.0(7.0)	&	4.5	&		\\
			322	&	40.93702(58)	&	0.037(4)	&	93.1(6.4)	&	4.9	&		&	359	&	26.21200(64)	&	0.033(4)	&	43.0(7.1)	&	4.4	&		\\
			323	&	33.49546(58)	&	0.037(4)	&	10.4(6.4)	&	4.9	&		&	360	&	28.15600(64)	&	0.033(4)	&	352.1(7.1)	&	4.4	&		\\
			324	&	13.92591(58)	&	0.037(4)	&	272.1(6.4)	&	4.9	&		&	361	&	32.02800(64)	&	0.033(4)	&	214.8(7.1)	&	4.4	&		\\
			325	&	13.95691(56)	&	0.038(4)	&	4.9(6.2)	&	5.1	&		&	362	&	23.12800(64)	&	0.033(4)	&	277.8(7.1)	&	4.4	&		\\
			326	&	12.96200(57)	&	0.038(4)	&	205.9(6.2)	&	5	&		&	363	&	23.19200(63)	&	0.034(4)	&	310.8(7.0)	&	4.5	&		\\
			327	&	3.99150(58)	&	0.036(4)	&	29.1(6.5)	&	4.8	&		&	364	&	1.20800(64)	&	0.033(4)	&	27.0(7.1)	&	4.4	&		\\
			328	&	25.99003(59)	&	0.036(4)	&	39.2(6.5)	&	4.8	&		&	365	&	0.80800(64)	&	0.033(4)	&	326.9(7.1)	&	4.4	&		\\
			329	&	32.17410(58)	&	0.036(4)	&	314.5(6.5)	&	4.8	&		&	366	&	15.02400(64)	&	0.033(4)	&	183.1(7.1)	&	4.4	&		\\
			330	&	0.65755(59)	&	0.036(4)	&	164.6(6.5)	&	4.8	&		&	367	&	23.75200(65)	&	0.033(4)	&	63.5(7.1)	&	4.4	&		\\
			331	&	37.96485(59)	&	0.036(4)	&	288.9(6.5)	&	4.8	&		&	368	&	30.63200(65)	&	0.033(4)	&	44.7(7.2)	&	4.4	&		\\
			332	&	40.99194(60)	&	0.036(4)	&	219.8(6.6)	&	4.8	&		&	369	&	22.22400(66)	&	0.032(4)	&	57.2(7.3)	&	4.3	&		\\
			333	&	16.60626(60)	&	0.035(4)	&	150.4(6.6)	&	4.7	&		&	370	&	0.16000(66)	&	0.032(4)	&	22.5(7.3)	&	4.3	&		\\
			334	&	18.21105(61)	&	0.035(4)	&	45.2(6.7)	&	4.7	&		&	371	&	25.36400(67)	&	0.032(4)	&	286.6(7.4)	&	4.2	&		\\
			335	&	36.66564(61)	&	0.035(4)	&	256.7(6.7)	&	4.6	&		&	372	&	32.10000(67)	&	0.032(4)	&	114.0(7.4)	&	4.2	&		\\
			336	&	38.35755(61)	&	0.035(4)	&	267.6(6.7)	&	4.6	&		&	373	&	0.07200(67)	&	0.032(4)	&	258.4(7.4)	&	4.2	&		\\
			337	&	32.37003(60)	&	0.036(4)	&	13.3(6.6)	&	4.7	&		&	374	&	30.66400(67)	&	0.032(4)	&	223.7(7.5)	&	4.2	&		\\
			338	&	36.54486(61)	&	0.035(4)	&	273.7(6.7)	&	4.6	&		&	375	&	30.15200(67)	&	0.032(4)	&	227.3(7.4)	&	4.2	&		\\
			339	&	38.19024(61)	&	0.035(4)	&	288.6(6.8)	&	4.6	&		&	376	&	36.58400(68)	&	0.031(4)	&	146.9(7.5)	&	4.2	&		\\
			\bottomrule																								
		\end{tabular}%}																												
	\end{table}																							\end{landscape}

\begin{landscape}	
	\begin{table}		
		\centering																											
		%\contcaption{   }																												
		%\scalebox{0.90}{																												
		\begin{tabular}{cc cc cl cc cc cl}																												
			\toprule																											
			$i$	&	$f_{\rm	i}$	&	$A$	&	$\Phi$	&	S/N	&	Combination				&	$i$	&	$f_{\rm	i}$	&	$A$	&	$\Phi$	&	S/N	&	Combination		\\
			&	(d$^{-1}$)	&	(mmag)	&	($^\circ$)	&	&						&		&	(d$^{-1}$)	&	(mmag)	&	($^\circ$)	&	&				\\
			\midrule																											
			377	&	21.98800(68)	&	0.031(4)	&	337.6(7.5)	&	4.2	&		&	386	&	27.12000(69)	&	0.031(4)	&	155.5(7.7)	&	4.1	&		\\
			378	&	0.60400(68)	&	0.031(4)	&	199.3(7.6)	&	4.1	&		&	387	&	26.86800(68)	&	0.031(4)	&	150.9(7.6)	&	4.1	&		\\
			379	&	24.72800(68)	&	0.031(4)	&	174.4(7.6)	&	4.1	&		&	388	&	26.76000(68)	&	0.031(4)	&	31.3(7.5)	&	4.2	&		\\
			380	&	11.54800(69)	&	0.031(4)	&	139.7(7.6)	&	4.1	&		&	389	&	40.55200(70)	&	0.030(4)	&	165.7(7.8)	&	4.0	&		\\
			381	&	12.04400(68)	&	0.031(4)	&	251.3(7.5)	&	4.2	&		&	390	&	33.04000(70)	&	0.030(4)	&	245.9(7.7)	&	4.0	&		\\
			382	&	31.75600(68)	&	0.031(4)	&	5.6(7.6)	&	4.1	&		&	391	&	26.03200(70)	&	0.030(4)	&	307.6(7.8)	&	4.0	&		\\
			383	&	22.86000(69)	&	0.031(4)	&	263.6(7.6)	&	4.1	&		&	392	&	30.45600(71)	&	0.030(4)	&	253.3(7.8)	&	4.0	&		\\
			384	&	24.38800(69)	&	0.031(4)	&	210.8(7.6)	&	4.1	&		&	393	&	27.76000(71)	&	0.030(4)	&	47.3(7.8)	&	4.0	&		\\
			385	&	38.91200(69)	&	0.031(4)	&	195.7(7.7)	&	4.1	&		&		&		&		&		&		&		\\
			\bottomrule																								
		\end{tabular}%}																												
	\end{table}																							\end{landscape}

%%%%%%%%%%%%%%%%%%%%%%%%%%%%%%%%%%%%%%%%%%
\reftitle{References}

\end{document}